\documentclass[twocolumn]{aastex631}
\usepackage{CJK}
\usepackage{amsbsy}
\usepackage{multirow}
\usepackage{bm}
\usepackage{ulem}

\usepackage{epsfig}
\usepackage{color}
\usepackage{graphicx}
\usepackage{longtable}
\usepackage{hyperref}

\def\kms{{\,\rm km\,s^{-1}}}
\def\msun{\,{\rm M}_\odot}
\def\msunh{\,{\rm h^{-1} M}_\odot}
\def\mpc{\,{\rm Mpc}}
\def\mpch{\,{\rm h^{-1}  Mpc}}

\def\e{\,{\boldsymbol{e_1}}}

\def\dzab{\,{\rm \Delta Z_{AB}}}
\def\dzabr{\,{\rm \Delta Z_{AB}^{random}}}
\def\dt{\,{\rm Z_{rms}/\Delta Z_{AB}}}

\received{XXX}
\revised{YYY}
\accepted{ZZZ}
\published{HHH}

\submitjournal{ApJ}

\shorttitle{filament spin}
\shortauthors{Peng Wang et al.}

\begin{document}
\begin{CJK*}{UTF8}{gbsn}
\title{Cosmic filament spin - II: filament spin and its impact on galaxy spin-filament alignment in a cosmological simulation}

\correspondingauthor{Peng Wang}
\email{pwang@shao.ac.cn}

\author[0000-0003-2504-3835]{Peng Wang* (王鹏)}
\affil{Shanghai Astronomical Observatory, Chinese Academy of Sciences, Nandan Road 80, Shanghai 200030, People's Republic of China}

\author[0009-0001-7527-4116]{Xiao-xiao Tang (唐潇潇)}
\affil{Shanghai Astronomical Observatory, Chinese Academy of Sciences, Nandan Road 80, Shanghai 200030, People's Republic of China}
\affil{University of Chinese Academy of Sciences, Beijing 100049, People's Republic of China}

\author{Hao-da Wang (汪昊达)}
\affil{Shanghai Astronomical Observatory, Chinese Academy of Sciences, Nandan Road 80, Shanghai 200030, People's Republic of China}
\affil{University of Chinese Academy of Sciences, Beijing 100049, People's Republic of China}

\author{Noam I. Libeskind}
\affil{Leibniz-Institut f\"ur Astrophysik Potsdam, An der Sternwarte 16, D-14482 Potsdam, Germany}

\author[0000-0002-5249-7018]{Elmo Tempel}
\affil{Tartu Observatory, University of Tartu, Observatooriumi 1, 61602 T\~oravere, Estonia}

\author{Wei Wang(王伟)}
\affil{Purple Mountain Observatory, Chinese Academy of Sciences, No.10 Yuan Hua Road, 210034 Nanjing, People's Republic of China.}
\affil{School of Astronomy and Space Science, University of Science and Technology of China, Hefei 230026, Anhui, People's Republic of China.}
\affil{Shanghai Astronomical Observatory, Chinese Academy of Sciences, Nandan Road 80, Shanghai 200030, People's Republic of China}

\author{Youcai Zhang (张友财)}
\affil{Shanghai Astronomical Observatory, Chinese Academy of Sciences, Nandan Road 80, Shanghai 200030, People's Republic of China}

\author[0000-0002-9891-338X]{Ming-Jie Sheng(盛明捷)}
\affil{Department of Astronomy, Xiamen University, Xiamen, Fujian 361005, People's Republic of China}

\author[0000-0001-5277-4882]{Hao-Ran Yu(于浩然)}
\affil{Department of Astronomy, Xiamen University, Xiamen, Fujian 361005, People's Republic of China}

\author[0000-0003-1132-8258]{Haojie Xu(许浩杰)}
\affil{Shanghai Astronomical Observatory, Chinese Academy of Sciences, Nandan Road 80, Shanghai 200030, People's Republic of China}


\begin{abstract}
Observational studies have reported that cosmic filaments on the megaparsec scale exhibit rotational motion. Subsequent simulation studies have shown qualitative agreement with these findings, but quantitative discrepancies remain due to differences in data and methods, which require verification. To address this issue, we adopt the same methodology as used in the observations to identify filament spin from the galaxy distribution constructed from a hydrodynamic simulation. Using the same approach to measure filament spin, we find that the simulation results closely match the observational findings, with only minor discrepancies arising from slight differences in the fraction of filaments classified as dynamically cold or hot based on their dynamic temperature. Additionally, an analysis of how filament spin affects the galaxy spin-filament correlation shows that filaments with strong spin signals and dynamically cold have a greater impact on the galaxy spin-filament correlation than those with weaker spin signals and dynamically hot filaments. These results not only provide further evidence that cosmic filaments exhibit spin, but also highlight the importance of this rotation in the acquisition of angular momentum by individual galaxies. Future studies exploring the influence of filament spin on galaxy spin may shed light on the physical origins of filaments and the angular momentum of galaxies.
\end{abstract}


\keywords{
    \href{http://astrothesaurus.org/uat/902}{Large-scale structure of the universe (902)};
    \href{http://astrothesaurus.org/uat/330}{Cosmic web (330)};
    \href{http://astrothesaurus.org/uat/2029}{Galaxy environments (2029)};
    \href{http://astrothesaurus.org/uat/1882}{Astrostatistics (1882)}}

\section{Introduction}
\label{sec:intro}
The large-scale structure of the Universe forms through gravitational instability, driven by initial perturbations in an otherwise homogeneous density field. Linear theory and the Zel'dovich approximation \citep{Zeldovich1970} predict that the matter distribution on megaparsec scales is not uniform but instead forms a connected network. In the latter half of the twentieth century, scientists such as \cite{Gregory1978, Joeveer1978, Shectman1996} began mapping the spatial organization of galaxies in the Universe. Detailed analyses of large-scale galaxy surveys, including the 2dF Galaxy Redshift Survey \citep{Colless2003}, the Sloan Digital Sky Survey \citep{Tegmark2004}, and the Two Micron All Sky Survey (2MASS) Redshift Survey \citep{Huchra2005}, have further confirmed these findings. These surveys reveal that galaxies are not distributed randomly but instead form a vast, intricate, web-like structure known as the cosmic web \citep{Bond1996Nature}. This cosmic web exhibits a multiscale, hierarchical structure \citep{Wangjie2020} that can be categorized into four main components: clusters (or knots), filaments, sheets (or walls), and voids. Clusters form in the densest regions, typically located at the intersections of filaments, and are fed by mass inflows along filament spines. Filaments, in turn, form at the intersections of walls, while walls arise along the boundaries of low-density voids.

Cosmic filaments are among the most significant structures in the cosmic web \citep{Bond1996Nature}. These quasilinear, extended topographical features of the galaxy distribution provide an environment that strongly influences galaxy formation and evolution, particularly through processes such as mass accretion \citep{Libeskind2014,ShiJingJing2015,KangWang2015} and galaxy environmental quenching \citep{Aragon2016galaxyquenching, winkel2021galaxyquenching}. Additionally, they are widely recognized for their crucial role in shaping and spinning dark matter halos \citep{2007MNRAS.381...41H, 2014MNRAS.440L..46A} and the galaxies they host \citep{TempelLibeskind2013}, especially in controlling rotational directions \citep{Zhang2009ApJ,Zhang2015ApJ}. The spins of galaxies and halos are correlated with the orientation of cosmic filaments, with this alignment depending on the mass \citep[see Table 1 in][and references within]{Wang2018ApJ.spin}. Low-mass galaxies and halos tend to align their spins with the host filament, while high-mass counterparts preferentially orient their spins perpendicular to the filament \citep{Wang2018ApJ.spin, ganeshaiah2021cosmic}. This transition in alignment occurs at a characteristic mass, estimated to range between $10^{11}\msunh$ and $10^{12}\msunh$, and has been linked to the properties of dark matter \citep{2020ApJ...902...22L, 2020ApJ...898L..27L, 2023ApJ...945...13M, 2024ApJ...966..100M}.The flip in halo spins from alignment to a perpendicular orientation is consistent with the ``two-stage'' evolution scenario \citep{WangKang2017,WangKang2018}, where the halo spin-filament correlation depends on whether mass accretion occurs along or perpendicular to the filament.

Previous studies focusing on the velocity field (or velocity shear field) around dark matter halos \citep{Codis2012, Libeskind2013, Laigle2015} have suggested that cosmic filaments may generate angular momentum themselves by providing a "rotational" environment for halos. \citet{Neyrinck2020} explored the possibility of filaments rotating by examining the motion of dark matter on the scale of cosmic filaments in a N-body simulation. More recently, \cite{Wang2021NatAS}, for the first time, provided observational evidence that cosmic filaments are spinning by comparing the redshift and blueshift of galaxies in filaments constructed from the Sloan Digital Sky Survey galaxy distribution \citep{Alam2015,Tempel2017catalogue}. A theoretical study \citep{xia2021filaspin} corroborated these findings using an N-body simulation, employing completely different methodologies for filament identification and spin definition. Although these two studies qualitatively agree, they exhibit some quantitative differences. This raises the question whether the $\rm \Lambda CDM$ model can accurately replicate observational results using the same methodology. To address this, we employ a state-of-the-art hydrodynamic cosmological simulation, utilizing the same filament finder as the observational study, Bisous \citep{Tempel2014bisous}, and applying an identical method to define spin.

This paper is organized as follows. Section \ref{sec:method} introduces the simulation data and explains the methodology used to measure the spin signal of the filaments. In Section \ref{sec:result}, we present the results of the filament spin signal obtained from the simulation and compare them with the observational data. Additionally, we examine the effect of filament spin on the correlation between galaxy spin and filament direction. Finally, in Section \ref{sec:sum_dis}, we summarize and discuss our findings.

\begin{figure}[!ht]
\plotone{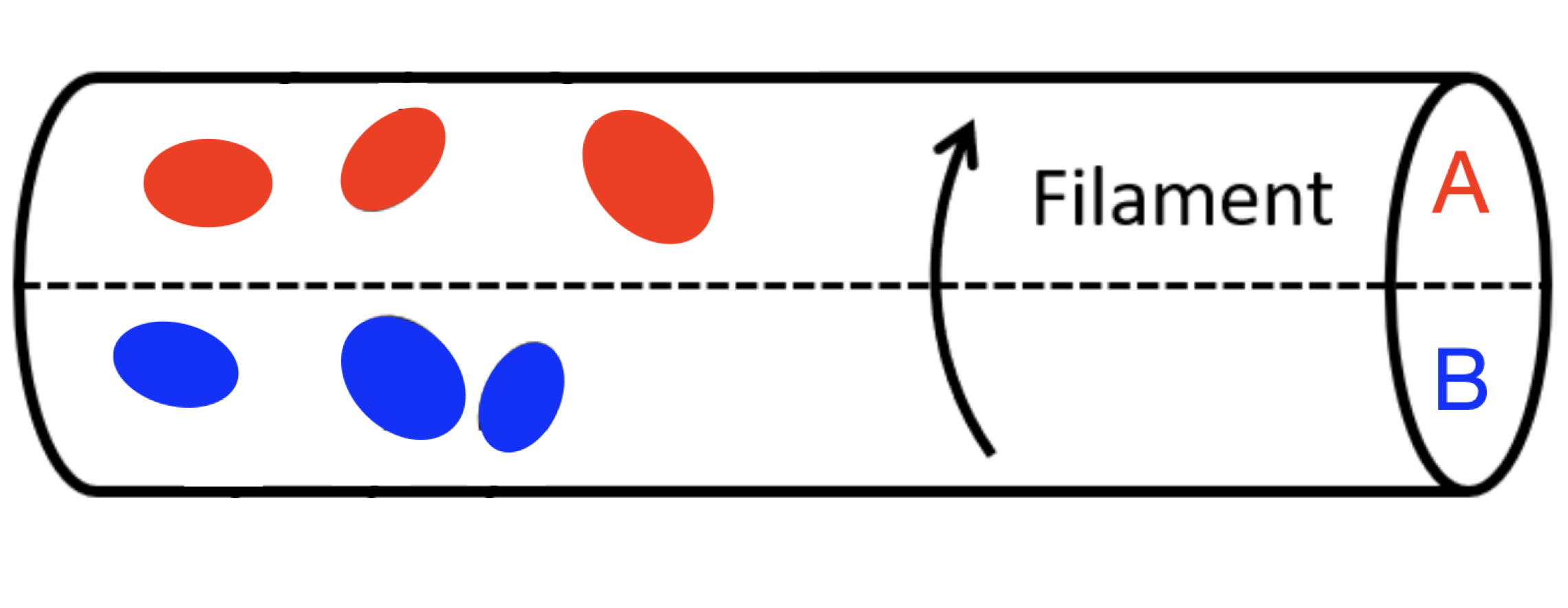}
\caption{A schematic cartoon diagram illustrating the filament spin calculation. The horizontal dashed line represents the filament spine, which divides the member galaxies into two regions (A and B) on either side of the spine. In this example, filament spine is perpendicular to the line-of-sight, there are three galaxies in each region, with a total of six galaxies. Only filaments with at least three galaxies per region and six galaxies in total are considered. Galaxies in region A are red-shifted relative to the filament, while those in region B are blue-shifted. From this, it can be inferred that the filament is rotating in the direction indicated by the arrows.}
\label{fig:f1}
\end{figure}

\section{Data and Methodology}
\label{sec:method}
We use the publicly available Illustris-TNG hydrodynamic cosmological simulation \citep{TNGpaper1,TNGpaper2,TNGpaper4,TNGpaper5}. Illustris-TNG adopts cosmological parameters from the Planck mission \citep{Planck2016}, including $\Omega_\Lambda = 0.6911$, $\Omega_{\rm m} = 0.3089$, and ${\rm H_0} = 100h \ \rm{km/s/Mpc}$, with $h = 0.6774$. The simulation suite consists of three box sizes: 50, 100, and 300 $\rm Mpc^3$, with each size available in both the N-body and hydrodynamic versions. For this study, we selected the TNG300-1 hydrodynamic simulation—the run with the largest box size—to ensure a statistically robust galaxy sample. This choice is motivated by our focus on the galaxy spin-filament spin correlation, which requires detailed stellar kinematics, and by our future goal of investigating interactions between the gaseous components of galaxies and those within filaments.
The simulation achieves a mass resolution of $\rm m_{\rm p} \sim 5.9 \times 10^7 M_{\odot}$ per particle. Dark matter halos are identified using the standard Friends-of-Friends (FoF) algorithm \citep{FoF1985}, while the SUBFIND algorithm \citep{sunfindSpringel2001,subfindDolag2009} is applied to each FoF group to identify gravitationally bound structures. The position of the most bound particle within each subhalo is used as the galaxy's position.

Considering the mass resolution and sample size, we selected galaxies with a stellar mass greater than $10^7 \msunh$. We obtained a total of 1,338,333 galaxies for the next step of the filament search.
The Bisous process, identical to the one used in the observational study of filament spin \citep{Wang2021NatAS}, was applied to the galaxy distribution to construct the filament catalogue. For detailed information on the Bisous algorithm, we refer the reader to \cite{Tempel2014bisous}, and for a comparison of the Bisous model with other cosmic web classification methods, to \cite{Libeskind2015bisous, Libeskind2018}.

To model the projection effects in the observations, we placed an imaginary observer at a distance of approximately 150 $\mpch$ from the center of the simulation box, corresponding to the average distance of galaxies in the observational study by \cite{Wang2021NatAS}. It is important to note that, in observations, the number density of galaxies decreases with distance for a given brightness limit because of observational constraints, whereas this effect does not occur in simulations. We will address this difference in Section~\ref{sec:sum_dis}. Using this set-up, we calculated the simulated redshift of galaxies as $\rm z = (H_0*d_{gal} + V_{los}) / c$, where the Hubble constant $\rm H_0=0.6774*100 \ km/s/Mpc$, $\rm d_{gal}$ is the proper distance from the galaxy to the imaginary observer, $\rm V_{los}$ is the line-of-sight peculiar velocity component, and $c$ is the speed of light.

In Fig.~\ref{fig:f1}, we illustrate the calculation of the spin of the filament. Each filament is modeled as a cylinder aligned with its spine and is viewed at an angle $\phi$ relative to the line of sight. The galaxies within each filament with a distance to the filament spine less than 2 $\mpc$ are divided into two regions, A and B. To ensure statistical reliability, each region must contain at least three galaxies, meaning that a filament must have a minimum of six galaxies in total. Applying this selection criterion, we obtained a sample of 29,190 filaments.

\begin{figure}[!ht]
\plotone{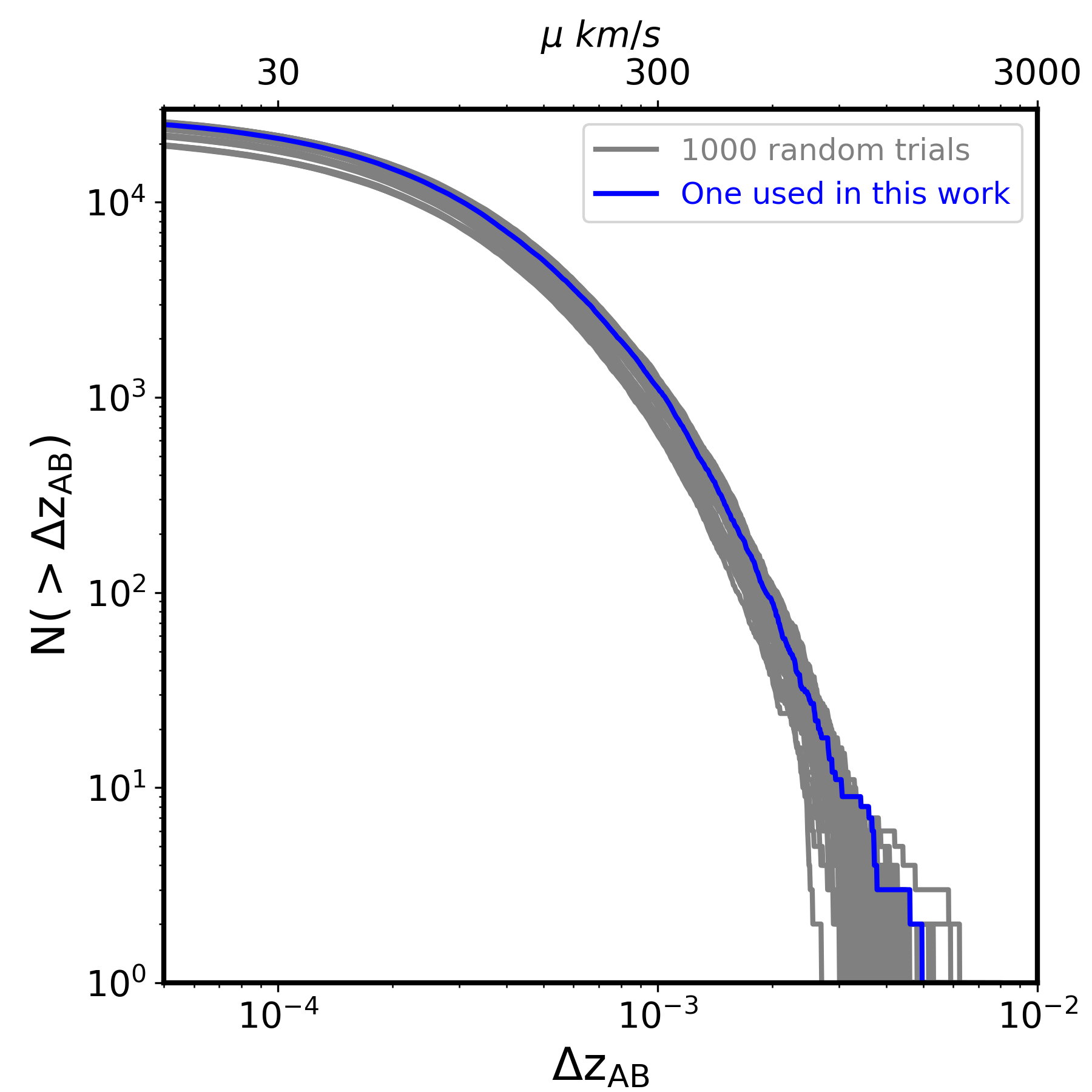}
\caption{The cumulative distribution of $\dzab$. The gray dashed lines represent 1,000 randomized trials, where the imaginary observer is placed randomly outside the simulation box while maintaining the same distance. The blue line represents a single randomized trial used in this work. The upper x-axis shows the rotation speed of the filament, calculated as $\rm \mu = c \times \dzab$. This indicates that the random placement of the observer does not significantly affect the measurement of filament spin, so we only randomly chose one trial in the following analyse.}
\label{fig:f2}
\end{figure}

The mean redshift, $z_{0}$, of a filament is calculated as the average redshift of all galaxies within it. Additionally, the mean redshifts of galaxies in regions A and B are denoted as $z_{A}$ and $z_{B}$, respectively. We hypothesize that if filaments rotate, there should be a noticeable velocity component perpendicular to the filament's axis. This component would act in opposite directions on either side of the filament's spine, with one side receding and the other approaching. To distinguish between the two regions based on their mean redshifts, we arbitrarily designate the region with the higher mean redshift as region A and the region with the lower mean redshift as region B. If the galaxies in region A are red-shifted relative to the filament ($z_{A} > z_{0}$), and those in region B are blue-shifted ($z_{B} < z_{0}$), the filament's spin can be inferred as indicated by the arrows in Fig.~\ref{fig:f1}. The difference in redshift between the two regions, $\rm \dzab = z_{A} - z_{B}$, serves as a proxy for the spin signal of the filament. The relative rotational speed is then calculated as $\mu = c \times \dzab$, where $c$ is the speed of light.

To classify filaments as dynamically hot or cold, we analyze the redshifts of all galaxies within each filament and calculate the root mean square of these values ($Z_{\rm rms}$). We compare $Z_{\rm rms}$ with $\dzab$ to make the distinction. Filaments with $Z_{\rm rms} > \dzab$ are classified as dynamically hot, while those with $Z_{\rm rms} < \dzab$ are classified as dynamically cold. Detecting rotation signals in dynamically hot filaments is inherently challenging, as $\rm \dzab$ values smaller than $Z_{\rm rms}$ lack physical significance. Consequently, significant rotation signals are not expected in dynamically hot filaments.

After determining $\dzab$ for each filament, it is necessary to assess the statistical significance of the measured value to ensure that the detected signal is not due to random fluctuations. To achieve this, we performed the following test: the redshifts of all member galaxies in a given filament were randomly shuffled while keeping their positions fixed. The same procedure for calculating $\dzab$ was then applied to the random trails. This process was repeated 10,000 times for each filament, generating a random distribution of $\dzabr$. The significance of the measured $\dzab$ can then be quantified as
\[
\rm s = \frac{\Delta Z_{AB} - \langle \Delta Z_{AB}^{\rm random} \rangle}{\sigma(\Delta Z_{AB}^{\rm random})}
\]
where $\rm \langle \dzabr \rangle$ and $\rm \sigma(\dzabr)$ are the mean and standard deviation of 10,000 random trials, respectively.

\begin{figure}[!htp]
\plotone{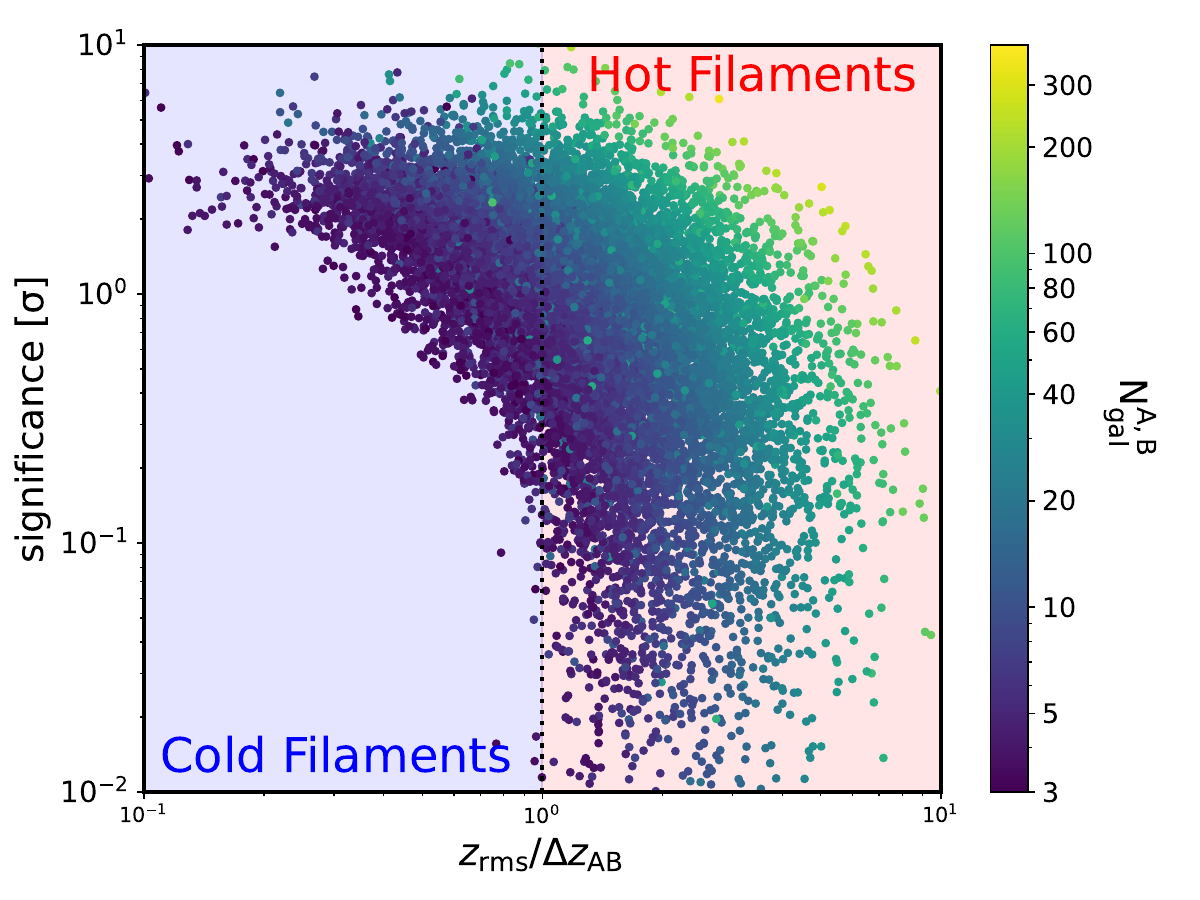}
\caption{The statistical significance of filament spin in the simulation is analyzed as a function of the filament's dynamic temperature. The dynamic temperature, $\dt$, of a given filament is defined as the ratio of the root mean square (rms) of the redshifts of the member galaxies, $\rm Z_{rms}$, to the redshift difference between those galaxies, $\dzab$. Filaments are classified as either "Cold" or "Hot" based on the criterion $\dt = 1$. Additionally, $\rm N_{gal}^{A,B}$ denotes the average number of galaxies on either side of the filament, as indicated by the color bar. }
\label{fig:f3}
\end{figure}


\section{Result}
\label{sec:result}

\subsection{filament spin signal}

To investigate the potential influence of the observer's placement on the filament spin measurement, we randomly selected an imaginary observer within our simulation. The distance between the observer and the center of the simulated box was fixed, but the observer's position was randomized. Following the procedure outlined in Sect.~\ref{sec:method}, we calculated the redshift of the galaxies, divided them into regions A and B of the filaments, and determined the redshift difference, $\dzab$. This process was repeated 1,000 times, and the results are presented in Fig.~\ref{fig:f2}. The gray lines represent the distributions of $\dzab$ of the 1,000 random tests, which are tightly clustered and show minimal variation. This indicates that the random placement of the observer does not significantly affect the measurement of filament spin. Given this consistency, we selected one of the random tests, shown as the solid blue line in Fig.~\ref{fig:f2}, for subsequent analysis. This result validates the robustness of our method against observer placement within the simulation.

\begin{figure*}[!ht]
\plotthree{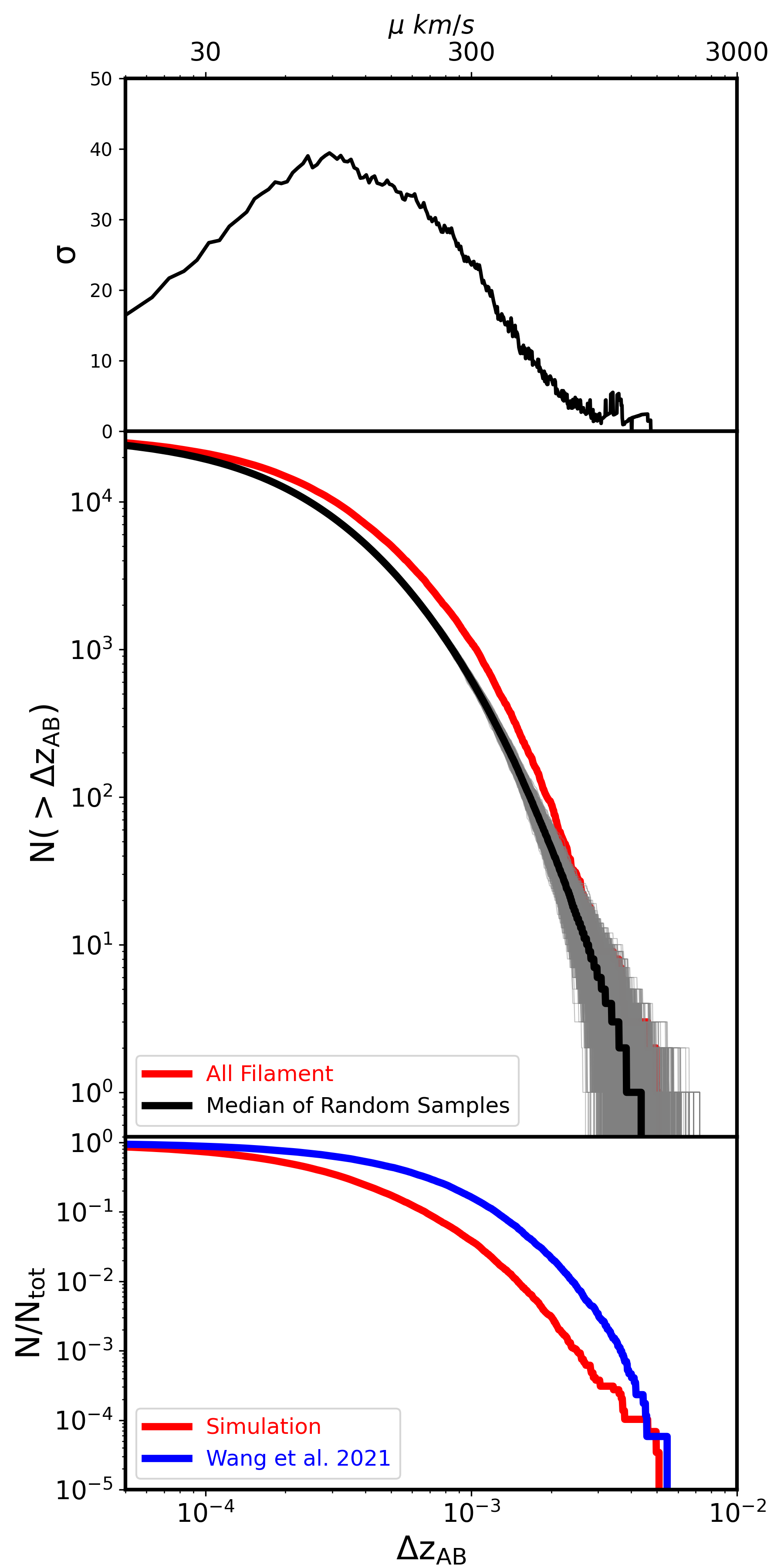}{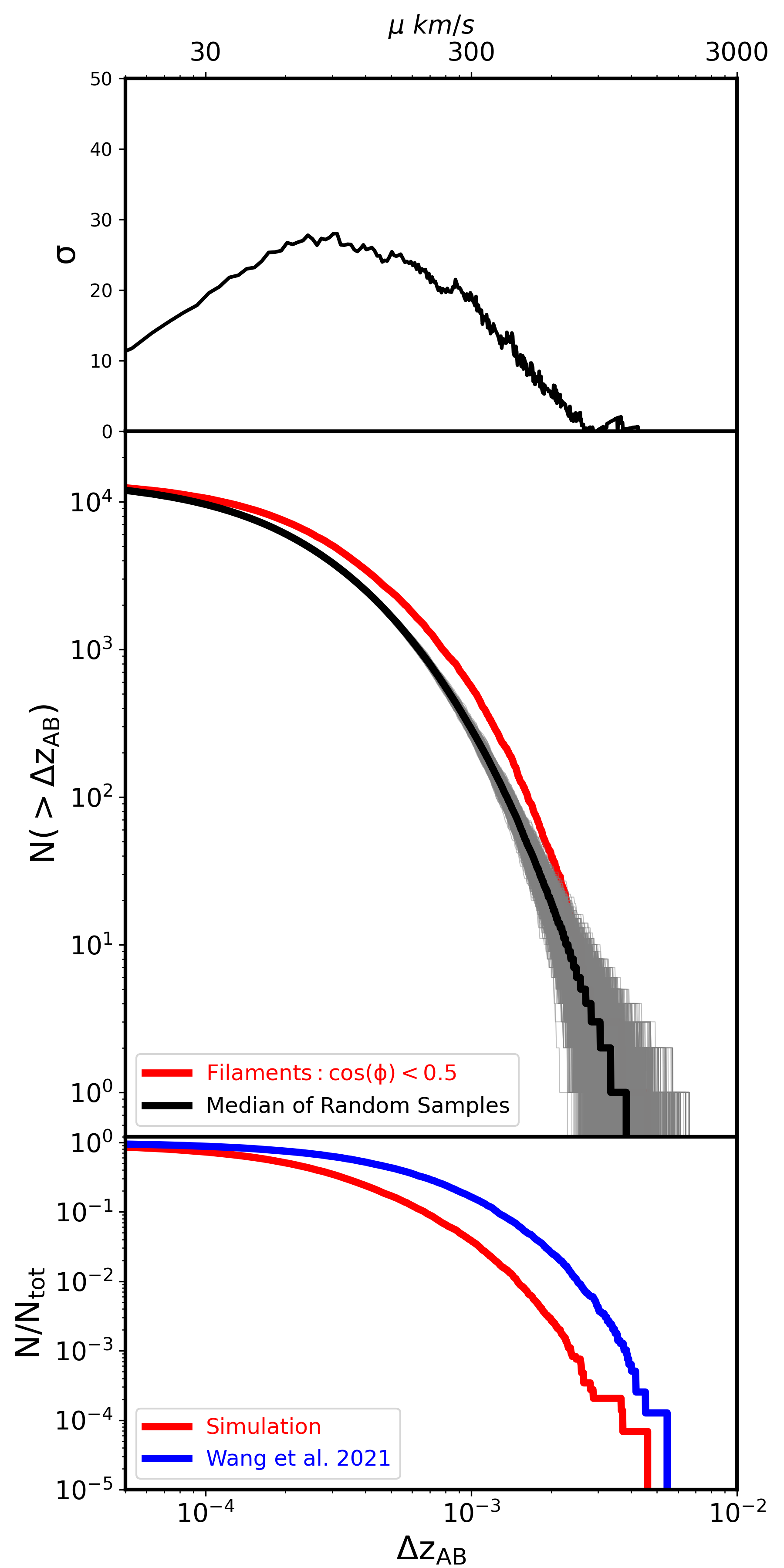}{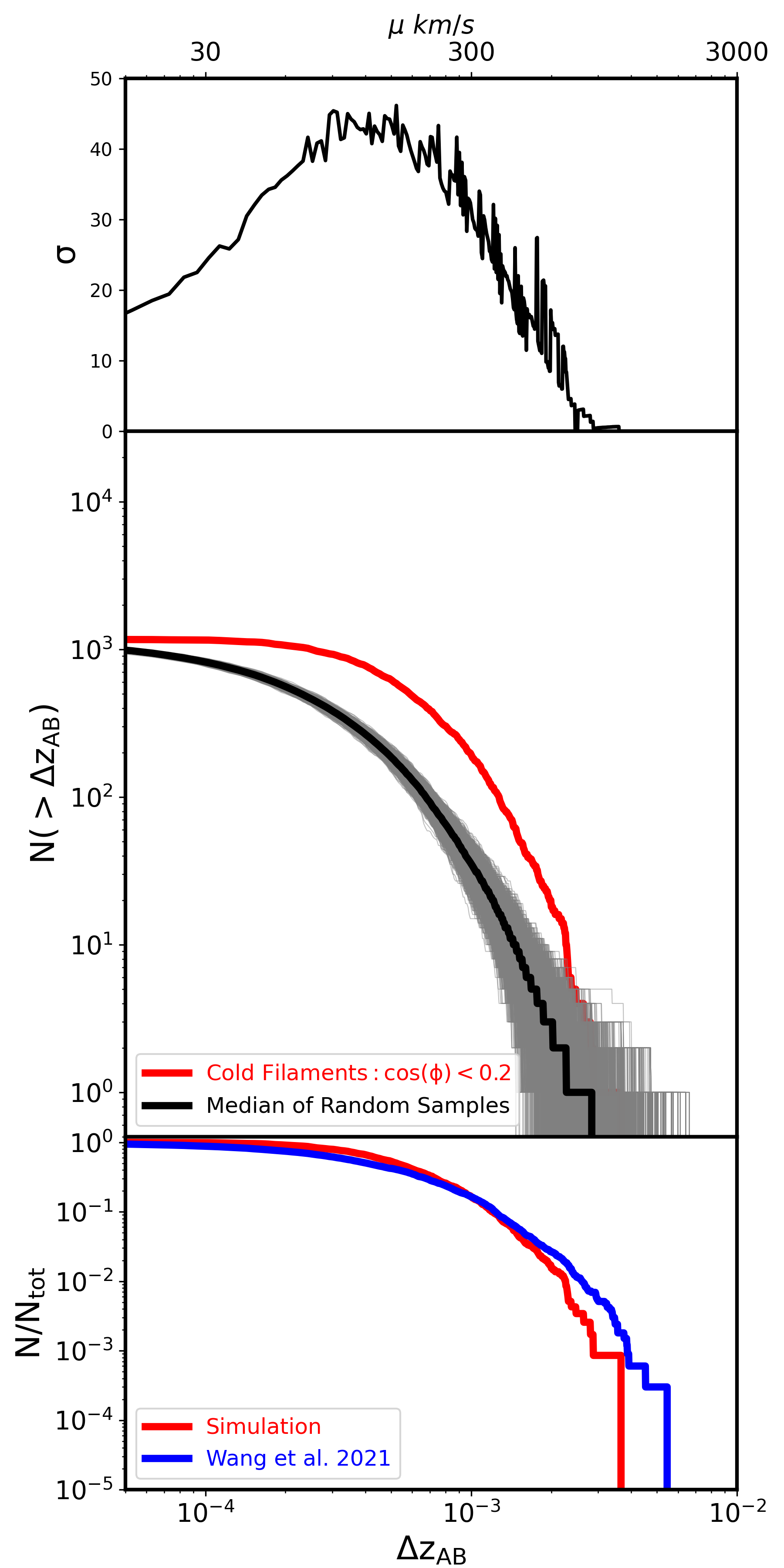}
\caption{The cumulative distribution of the redshift difference, $\dzab$, between galaxies in region A and region B within filaments is analyzed. From left to right, each column corresponds to: all filaments without considering the viewing angle (left column), filaments whose spines subtend an angle $\rm \cos(\phi) < 0.5$ relative to the line of sight (middle column), and dynamically cold filaments whose spines subtend an angle $\rm \cos(\phi) < 0.2$ and satisfy $\dt < 1$ (right column). The solid red lines in the middle rows represent the distribution of the measured filament signal, while the 10,000 gray lines correspond to randomized samples. The black solid lines indicate the median values of the random samples. The top panels display the sample-to-sample distance between the real filaments in the simulation and the random samples, expressed in units of the standard deviation of the random trials. The upper x-axes show the rotation speed of the filaments, calculated as $\rm \mu = c \times \dzab$. The bottom panels compare each subsample from the simulation with observational data \citep{Wang2021NatAS}.}
\label{fig:f4}
\end{figure*}

In Fig.~\ref{fig:f3}, we present the statistical significance of the measured $\dzab$ for the filaments, expressed in terms of the number of standard deviations ($\sigma$) by which it differs from the random sample, as a function of the dynamic temperature of the filament $\dt$. This relationship indicates that as the dynamic temperature of a filament increases, the statistical significance of the $\dzab$ signal (a proxy for the filament's spin) decreases. In other words, filaments with higher dynamic temperatures are less likely to exhibit a significant $\dzab$ signal, suggesting a weaker or negligible spin. In contrast, colder filaments (lower dynamic temperature) show higher $\sigma$ values, meaning that their redshift differences are more inconsistent with the random sample. This implies a stronger signal for filament spin in colder filaments. This trend highlights the connection between the dynamic temperature of a filament and the detectability of its spin signal.

The mean number of member galaxies in regions A and B is represented by color, ranging from dark blue (smallest value of 3, noting that 6 is the minimum number of galaxies per filament) to light yellow (several hundred galaxies), as shown in the color bar in Fig.~\ref{fig:f3}. Compared to observations, filaments in simulations contain significantly more galaxies. Observations show a maximum of $\sim$90 galaxies per filament \citep[see Fig.~1 in][]{Wang2021NatAS}, while simulations exhibit slightly more than 300 galaxies per filament. Importantly, the ratio of galaxies in region A to region B remains nearly equivalent. We found that filaments with a larger number of member galaxies exhibit a more inconsistent (higher $\sigma$) redshift difference, $\dzab$, compared to random samples, resulting in a more significant spin signal.

A natural question arises: why does a greater number of member galaxies lead to a more pronounced rotation signal? Previous studies, such as \cite{WangKang2018,Neyrinck2020,Wang2021NatAS}, proposed that galaxies orbit filaments in helical motion. As galaxies move along these orbits, their velocities transition from being perpendicular to aligning with the filament's direction. If filaments are indeed spinning, then a filament containing a larger number of galaxies provides a more accurate representation of its rotation dynamics, enhancing the observed spin signal.

However, it is important to note that not all filaments exhibit spin. Similarly to galaxies being classified as spiral or elliptical based on the motion patterns of their stars, filaments can also be categorized by their dynamic temperatures. Specifically, we classified filaments as cold if $\dt < 1$ and hot if $\dt > 1$, as indicated by the blue and red regions, respectively, in Fig.~\ref{fig:f3}. Our analysis shows that the statistical significance of filament spin decreases with increasing filament dynamic temperature. 
This suggests that dynamically cold filaments are more likely to exhibit spin.
When we account for the number of galaxies in a filament, we find an additional trend: for filaments with a given spin significance, those with more member galaxies tend to be dynamically hotter. 

In Fig.~\ref{fig:f4}, we present the cumulative distribution of the filament spin signal, $\dzab$, which corresponds to the redshift difference of galaxies between regions A and B. Observational studies \cite{Wang2021NatAS} and \cite{tangxx2025} have shown that the spin signal of the filament depends on the viewing angle of the filament spine relative to the line of sight. To explore this, we examined the dependence of $\dzab$ on the viewing angle. 
Fig.~\ref{fig:f4} shows three cases: Left column: All filaments, regardless of the viewing angle.
Middle column: Filaments with spine angles relative to the line of sight such that $\rm \cos(\phi) < 0.5$.
Right column: Dynamically cold ($\rm \dt < 1$) filaments with viewing angle $\rm \cos(\phi) < 0.2$.

In the central rows of Fig.~\ref{fig:f4}, the solid red lines represent the distribution of the measured filament spin signals, while the 10,000 gray lines denote random samples. For random trials, we kept the position of galaxies fixed, but randomly shuffled their redshifts (see Section.~\ref{sec:method} for more details). The median values of the random samples are represented by the black solid lines. The red lines and black lines remain distinct and do not intersect, except when $\dzab$ is very small or very large. The separation between the red and black lines increases as the cosine value of the viewing angle decreases. This effect becomes even more pronounced when the condition of dynamically cold filaments is applied. The dependency of this separation on $\dzab$ is clearly illustrated in the upper panels. This analysis highlights the influence of the viewing angle on the observed spin signal and confirms that dynamically cold filaments with smaller angles relative to the line of sight exhibit stronger spin signals.

The upper panels of Fig.~\ref{fig:f4} quantify the separation between filaments in the simulation and random samples, normalized by the standard deviation of the random trials, as a function of the filament spin signal, $\dzab$. The upper x-axes show the rotation speed of the filaments, calculated using $\rm \mu = c \times \dzab$, where $c$ is the speed of light. The peak values of the curves range from $30\sigma$ to $50\sigma$, indicating a strong cumulative statistical signal for the spin of the filaments.

The lower panels of Fig.~\ref{fig:f4} provide a direct comparison between sub-samples of filaments in the simulation and those from observations \citep{Wang2021NatAS}. For the first two subsamples, the observational signal (depicted as solid blue lines) is higher than the simulation signal for most $\dzab$ values. In contrast, for the final subsample, the observational and simulation signals nearly overlap. This discrepancy arises from differences in the ratio of cold to hot filaments between observations and simulations. When both the viewing angle and the dynamically cold filaments are applied, the signals in the observations and simulations become consistent.

\begin{figure}[!t]
\plotone{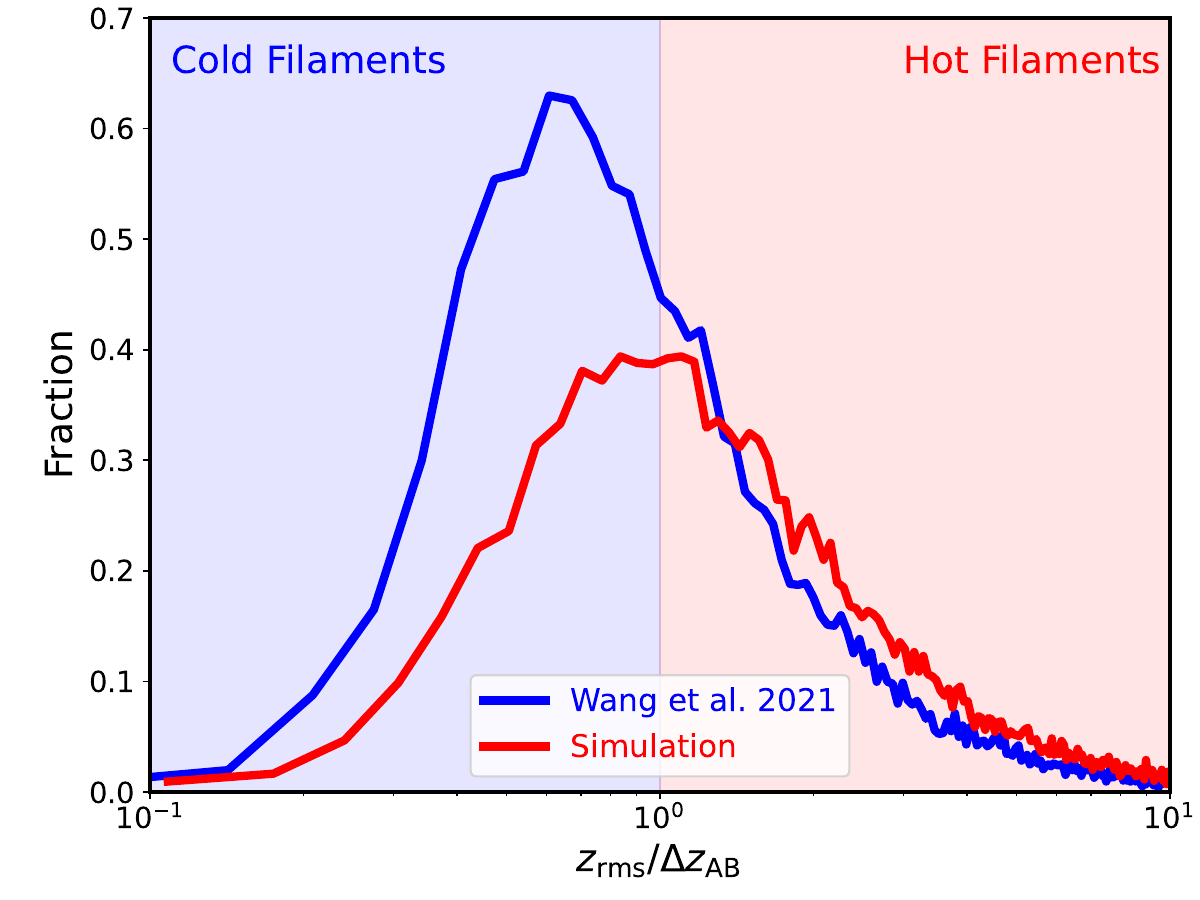}
\caption{The distribution of filament dynamic temperature, $\dt$, is shown for both the simulation (solid red line) and observation (solid blue line). Filaments are categorized as either ``Cold" or ``Hot" based on the criterion $\dt = 1$.}
\label{fig:f5}
\end{figure}

In Fig.~\ref{fig:f5}, we present the dynamic temperature distribution of filaments, comparing observations with simulations. The blue line peaks at $\dt < 1$, while the red line peaks at $\dt \sim 1$, indicating more number of hot filaments in the simulation compared to observations. 
We note the discrepancy in the fraction of dynamically cold/hot filaments between observation and simulation, which may arise from several factors. In observations, filaments are primarily defined by the distribution of galaxies, and observational catalogs are inherently biased toward filaments populated by brighter or more massive galaxies. In contrast, simulations identify filaments directly from galaxy populations with a specified mass threshold. Additionally, systematic errors in the observational data may also contribute to this discrepancy.

\begin{figure}[!ht]
\plotone{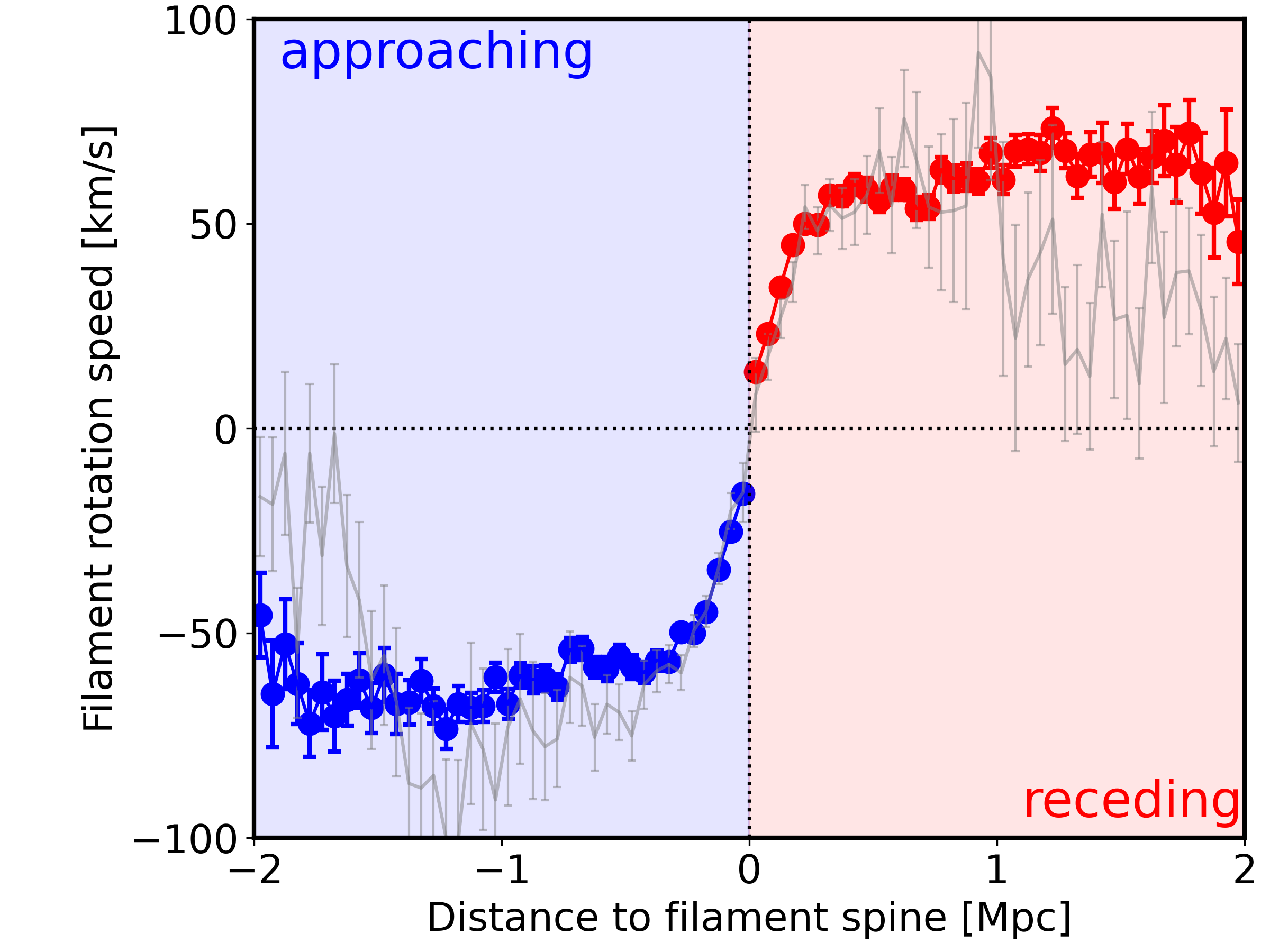}
\caption{The stacked rotation curve of galaxies in filaments is shown, where the rotation speed, $c \times \Delta z$, is plotted as a function of the distance between galaxies and the filament spine. Here, $\Delta z$ represents the redshift difference between all galaxies within a given distance bin and the mean redshift of the filament. The red and blue dots with error bars indicate results from the simulation, while the gray line corresponds to observational data \citep{Wang2021NatAS}. Galaxies in regions moving away from the observer are shown in red, with their distances from the filament spine and rotation speeds assigned positive values. Conversely, galaxies in regions moving toward the observer are shown in blue, with their distances and speeds assigned negative values.}
\label{fig:f6}
\end{figure}

The above analysis examines the spin signal of individual filaments (Fig.\ref{fig:f3}) and the stacked signal of all filaments (Fig.\ref{fig:f4}). However, each galaxy within a given filament has a specific distance from the filament spine, allowing us to measure the stacked rotation signal of galaxies at a given distance. In Fig.~\ref{fig:f6}, we present the stacked rotation curve of galaxies in filaments as a function of their distance from the filament spine. The simulation results are shown as red and blue points with error bars, while the gray line with error bars represents the observational results from \cite{Wang2021NatAS}. Galaxies receding from the observer are marked in red, and their distances from the filament spine are treated as positive. Conversely, galaxies approaching the observer are marked in blue, with their distances considered negative. The same approach is applied to define the positive or negative values of the rotation speed.

As seen in Fig.~\ref{fig:f6}, the rotation curves from the simulation (red and blue lines) roughly agree with those from the observations (gray line). The error bars are smaller in the simulation because of the larger number of galaxies and filaments. In the inner regions of filaments, specifically at distances less than 1 Mpc, the observational curves and simulations align closely. However, this agreement weakens as the distance increases. In observations, the rotation curves drop more sharply at distances greater than 1 Mpc. Although the margins of error are significant, a clear trend toward zero is evident. In contrast, the rotation curve in the simulation shows minimal reduction, with the rotation curve gradually decreasing beyond $1.8 \mpc$, reaching approximately $50 \kms$ at 2 Mpc.

The discrepancies of the rotation curves between the simulation and the observation can be attributed to the following factors:

First, the discrepancy may stem from the differing proportions of cold and hot filaments.  As shown in Fig.~\ref{fig:f5},  an approximately equal count of hot and cold filaments is present in the simulation sample, in contrast to the observation sample, which shows more number of cold filaments. Further support is demonstrated in the lower right panel of Fig.~\ref{fig:f4}, which shows that the simulation results align closely with the observational findings when only cold filaments are considered.

Secondly, as shown by the gray line in Fig.~\ref{fig:f6}, although the rotation curve in the observational data decreases at distances greater than 1 Mpc, the uncertainty increases significantly within the 1$\sim$2 Mpc interval. This is primarily due to the limited number of galaxies in the observational filaments, a situation that is not reflected in the simulations. In the simulations, while the number of galaxies generally decreases with increasing distance from the filament spine \citep{Daniela2020, Daniela2024, 2024MNRAS.532.4604W}, a significant population of low-mass galaxies is still included. This differs from the observational data, where low-mass galaxies are often undetected because of observational limitations. Consequently, the simulations produce more precise rotation profiles, particularly in the outer filament regions, with smaller errors compared to the observational sample.

In addition, a test was conducted using a higher mass threshold for galaxy selection within the simulation and focusing on dynamically cold filaments. Although the resulting rotation curves showed greater consistency with the observations, slight differences remain. \cite{xia2021filaspin} using filaments from an N-body simulation, conducted a similar study on filament spin and found that the rotation curve reached approximately 70 $\kms$ at 2 $\mpc$. This result is more consistent with our finding (50 $\kms$), although discrepancies still persist, probably due to differences in the data and methodologies used.

The third point is that, in observations, a phenomenon similar to the asymmetric drift effect seen in the rotation profiles of disk galaxies \citep{bershady2024asymmetric} may be present. This effect causes the outer rotation velocity to decrease more rapidly than expected. We suspect that a similar effect occurs when the rotation profiles of filaments are measured in observational data, but we need more studies to confirm this.

\subsection{effect of filament spin on galaxy spin}

Over the past decades, numerous researchers have highlighted, through both observation \citep{TempelLibeskind2013, Zhang2015ApJ, Pahwa2016} and simulation \citep{Codis2012, trowland2012cosmic, dubois2014dancing, WangKang2017, WangKang2018, ganeshaiah2018cosmic, ganeshaiah2019cosmic, ganeshaiah2021cosmic} studies, that the spins of galaxies are influenced by their host filament. In this work, we revisit this effect while considering the rotation of the filaments.

\begin{figure*}[!ht]
\plottwo{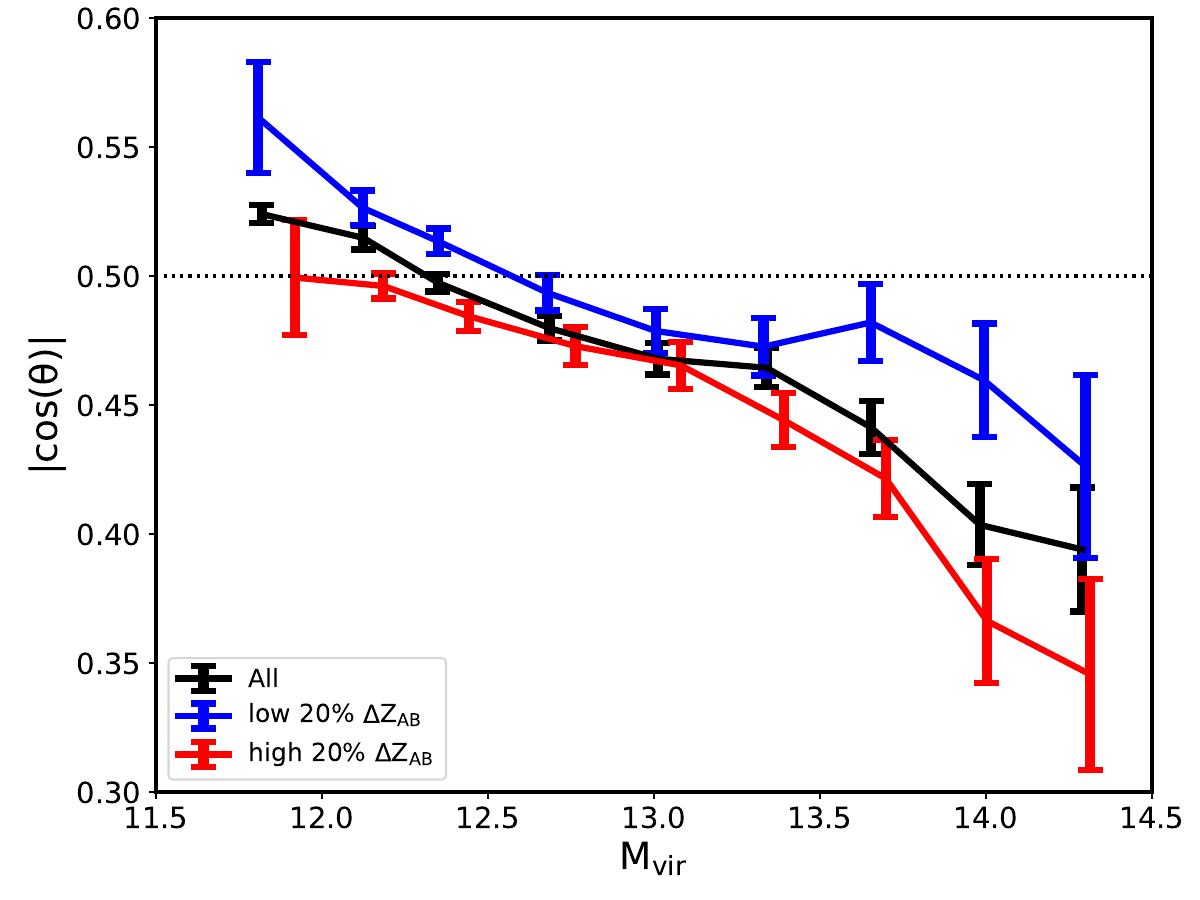}{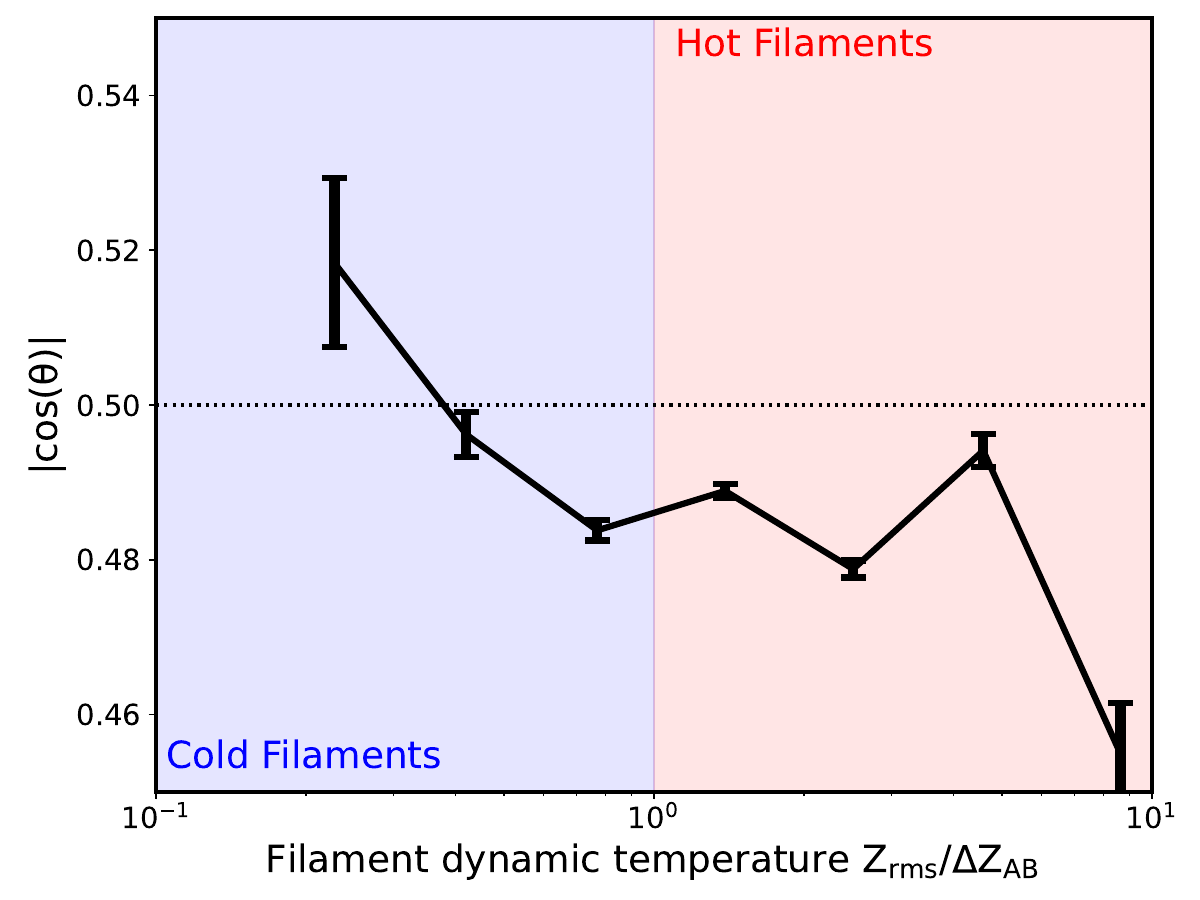}
\caption{Left panel: The mass dependence of $|\cos(\theta)|$, where $\theta$ is the angle between the galaxy spin and the orientation of its host filament. The black solid line represents all samples, while the red and blue solid lines correspond to galaxies in filaments with the highest and lowest 20\% of spin signal values, respectively. The dotted line represents the case where galaxy spins are randomly distributed relative to their host filaments. Values of $|\cos(\theta)| > 0.5$ indicate a preferential alignment between galaxy spins and filament directions, while $|\cos(\theta)| < 0.5$ suggests a perpendicular alignment. Right panel: The $|\cos(\theta)|$ as a function of filament dynamic temperature. Filaments are classified as "cold" or "hot" based on their dynamic temperature, with the division marked by the vertical dotted line.}
\label{fig:f7}
\end{figure*}

In the left panel of Fig.~\ref{fig:f7}, we present the distribution of the absolute value of $\cos(\theta)$ for all galaxies in our sample, where $\theta$ is the angle between the galaxy spin and the direction of the filament spine. . This distribution is represented by the solid black line with error bars. Statistically, when the orientation of a galaxy's spin is randomly distributed relative to the filament, the expected value of $|\cos(\theta)|$ is 0.5. If $|\cos(\theta)|$ exceeds 0.5, it indicates an alignment between the galaxy spin and the filament. Conversely, if $|\cos(\theta)|$ is below 0.5, it suggests that the galaxy spin is preferentially perpendicular to the filament. Consistent with previous studies \citep[e.g.,]{Wang2018ApJ.spin}, we confirm the trend that the value of $|\cos(\theta)|$ decreases with the virial mass of the dark matter halo hosting the galaxies, indicating a parallel trend for low-mass halos and a perpendicular trend for high-mass halos. The transition occurs at a mass around $10^{12.3}\msun$.

To examine the effect of filament rotation on the galaxy spin-filament correlation, we divided the filaments into two sub-samples: the top 20\% and the bottom 20\% based on their $\dzab$ values (how fast the filament spin), represented by the red and blue solid lines, respectively. It can be observed that the red solid line is systematically lower than the blue solid line, indicating that the spins of galaxies tend to be more perpendicular to the direction of the filaments with larger rotation velocities. In contrast, galaxies in filaments with smaller rotation velocities exhibit a tendency to align more parallel to the filament direction.

In the right panel of Fig.~\ref{fig:f7}, we present $|\cos(\theta)|$ as a function of the dynamic temperature of the filament $\dt$. It is evident that the galaxy spin-filament correlation depends on $\dt$. As $\dt$ increases, $|\cos(\theta)|$ decreases, transitioning from a parallel trend to a perpendicular trend. This occurs because galaxies in dynamically hot filaments tend to have higher masses compared to those in dynamically cold filaments.

According to the two-phase model of the formation and evolution of the spin-filament correlation proposed by \cite{WangKang2017} and \cite{WangKang2018}, the correlation depends on the timescale over which a galaxy resides in its surrounding environment, that is, the duration the galaxy spends in a wall or filament. The longer a galaxy resides in the wall (filament), the more likely it is to align with (become perpendicular to) the filament direction. Since filaments form as a result of the collapse of walls along the secondary direction of matter collapse, it is important to note that filaments undergo their own evolution, characterized by continuous changes in their density distribution \citep{Daniela2020,Daniela2022,Daniela2024} and radius \citep{2024MNRAS.532.4604W}. During the collapse of the wall into the filament, the radius of the filament gradually decreases \citep{2024MNRAS.532.4604W}, while its density increases, and its rotation velocity $\dzab$ can also increase correspondingly. These changes make the filament more constraining for the galaxies within it. In other words, such filaments are established earlier, providing galaxies within them more time to be influenced by the filament. Consequently, these galaxies acquire an increased perpendicular angular momentum by accrete more matter (subhalo, satellties) along the filament direction.

\section{Summary and Discussion}\label{sec:sum_dis}
Motivated by the observational findings on filament spin reported by \cite{Wang2021NatAS}, we investigate the filament spin signal in a hydrodynamic simulation based on the $\rm \Lambda CDM$ cosmological model to assess the possibility of reproducing this phenomenon. Additionally, we revisit the galaxy spin–filament correlation, incorporating the direction of filament spin into our analysis.

In this study, we used the Bisous model \citep{Tempel2014bisous} to identify filaments in the galaxy distribution selected from the hydrodynamic simulation TNG300-1 \citep{TNGpaper1,TNGpaper2,TNGpaper4,TNGpaper5}. The primary results of our analysis are as follows.

\begin{itemize}
    \item The observed filament spin signal is qualitatively reproduced in the simulation.
    \item Quantitative evaluation reveals discrepancies in the filament spin signals, which arise from variations across the filament and galaxy samples. However, these discrepancies are within acceptable limits.
    \item Incorporating filament spin, we investigate the correlation between the spins of galaxies and filaments. Our findings indicate that filament spin influences this correlation.
\end{itemize}

Using a method nearly identical to that employed in the observational study by \cite{Wang2021NatAS} to measure the filament spin signal, our conclusions about the filament spin signal are broadly consistent with previous results \cite{Wang2021NatAS, xia2021filaspin}.

Some differences, revealed through quantitative comparative analysis, have also been discussed in the work of \cite{tangxx2025}. They noted that while qualitative conclusions can be obtained using different observational galaxy data and filament identification algorithms, the differences in quantitative analyzes are significant.

It is important to note that we did not use the mock method or directly calculate the angular momentum of the filaments from the simulation \citep[as done by][ ]{Sheng2022PhRvD}. Therefore, on the one hand, it is crucial to employ the mock method to fully reproduce the observed results from numerical simulations, particularly for quantitative analyses. On the other hand, directly investigating the angular momentum of filaments and their formation and evolution by analyzing the velocity field around the filaments, specifically the motion of galaxies or dark matter particles, forms an essential part of our future plans.

Our current analysis centers on the global spin signal of filaments, as a systematic investigation of its spatial variation along the filamentary direction has not yet been conducted. We emphasize that this phenomenon is highly sensitive to the filament detection algorithm, given that divergent methodological frameworks for defining and reconstructing filamentary structures (e.g., density thresholds, connectivity criteria) may introduce significant systematic biases. Nevertheless, probing potential torsional modes or spatially dependent spin patterns through segment-wise analysis of filaments represents a promising avenue for further research. Should such variations be robustly identified, their physical implications (e.g., angular momentum transfer mechanisms, dark matter-filament coupling) will be rigorously examined. These questions will be prioritized in future studies leveraging higher-resolution simulations and multi-algorithm cross-validation.

The study of filament rotation in cosmology has a wide range of applications. For example, \cite{2022PhLB..83337298A} used observed filament rotation curves to constrain the mass range of ultralight dark matter, establishing both upper and lower bounds. Furthermore, \cite{2023MNRAS.519.1171Z} investigated the kinetic Sunyaev-Zeldovich (KSZ) effect by analyzing aggregated signals from multiple filaments.

In the correlation between filaments and the angular momentum of galaxies, it is important to note that, particularly in observations, determining both the direction of the filament and the galaxy's spin is often challenging. For spiral galaxies, the orientation of spin vectors is typically inferred using right ascension (RA), declination (Dec), and the inclination angle derived from the projected axis ratio, as noted in studies by \cite{Trujillo2006, Lee2007, Varela2012}. Conversely, for elliptical galaxies, it is common to deduce the projected spin orientation by assuming that the minor axis aligns with the spin direction, as described by \cite{TempelLibeskind2013}. However, this assumption is often inaccurate because of a misalignment between the true spin direction and the minor axis orientation, leading to significant measurement discrepancies.

Additionally, the alignment of galaxy spins with cosmic filaments, whether aligned in the same direction or opposite, carries distinct physical implications and origins. However, this aspect has often been overlooked in previous studies by checking the absolute value of the angle between the galaxy spin and the filament direction, that is, $\rm |\cos(\theta)|$. Although the spin of filaments can provide insight into the direction of the filament, its physical interpretation requires further exploration. Similarly, the spin of galaxies can be better constrained with more precise observations, such as those provided by MaNGA \citep{MaNGA2015Bundy}.

In recent work, we investigated the correlation between the spin direction of the filament and the spin directions of galaxies indicated by the stellar components and the H$\alpha$ line in MaNGA. By removing the absolute value of the angle between galaxy spin and filament direction, we found that the spin direction of low-mass disk galaxies tends to be opposite to that of the filamentary structure. However, these findings require further verification through numerical simulations. This calls for more detailed studies, which we aim to pursue in future work.

\begin{acknowledgments}

The authors thank anonymous referees for comments that substantially improved the manuscript.  We acknowledge stimulating discussions with Fangzhou Jiang.
PW, XXT, WHD acknowledge the financial support from the NSFC (No.12473009), and also sponsored by Shanghai Rising-Star Program (No.24QA2711100). YCZ acknowledges the financial support from the NSFC (No.12273088). ET acknowledges funding from the HTM (grant TK202), ETAg (grant PRG1006) and the EU Horizon Europe (EXCOSM, grant No. 101159513).

\end{acknowledgments}

\bibliography{main}{}

\begin{thebibliography}{}
\expandafter\ifx\csname natexlab\endcsname\relax\def\natexlab#1{#1}\fi
\providecommand{\url}[1]{\href{#1}{#1}}
\providecommand{\dodoi}[1]{doi:~\href{http://doi.org/#1}{\nolinkurl{#1}}}
\providecommand{\doeprint}[1]{\href{http://ascl.net/#1}{\nolinkurl{http://ascl.net/#1}}}
\providecommand{\doarXiv}[1]{\href{https://arxiv.org/abs/#1}{\nolinkurl{https://arxiv.org/abs/#1}}}

\bibitem[{{Alam} {et~al.}(2015){Alam}, {Albareti}, {Allende Prieto}, {Anders},
  {Anderson}, {Anderton}, {Andrews}, {Armengaud}, {Aubourg}, {Bailey}, {Basu},
  {Bautista}, {Beaton}, {Beers}, {Bender}, {Berlind}, {Beutler}, {Bhardwaj},
  {Bird}, {Bizyaev}, {Blake}, {Blanton}, {Blomqvist}, {Bochanski}, {Bolton},
  {Bovy}, {Shelden Bradley}, {Brandt}, {Brauer}, {Brinkmann}, {Brown},
  {Brownstein}, {Burden}, {Burtin}, {Busca}, {Cai}, {Capozzi}, {Carnero
  Rosell}, {Carr}, {Carrera}, {Chambers}, {Chaplin}, {Chen}, {Chiappini},
  {Chojnowski}, {Chuang}, {Clerc}, {Comparat}, {Covey}, {Croft}, {Cuesta},
  {Cunha}, {da Costa}, {Da Rio}, {Davenport}, {Dawson}, {De Lee}, {Delubac},
  {Deshpande}, {Dhital}, {Dutra-Ferreira}, {Dwelly}, {Ealet}, {Ebelke},
  {Edmondson}, {Eisenstein}, {Ellsworth}, {Elsworth}, {Epstein}, {Eracleous},
  {Escoffier}, {Esposito}, {Evans}, {Fan}, {Fern{\'a}ndez-Alvar}, {Feuillet},
  {Filiz Ak}, {Finley}, {Finoguenov}, {Flaherty}, {Fleming}, {Font-Ribera},
  {Foster}, {Frinchaboy}, {Galbraith-Frew}, {Garc{\'\i}a},
  {Garc{\'\i}a-Hern{\'a}ndez}, {Garc{\'\i}a P{\'e}rez}, {Gaulme}, {Ge},
  {G{\'e}nova-Santos}, {Georgakakis}, {Ghezzi}, {Gillespie}, {Girardi},
  {Goddard}, {Gontcho}, {Gonz{\'a}lez Hern{\'a}ndez}, {Grebel}, {Green},
  {Grieb}, {Grieves}, {Gunn}, {Guo}, {Harding}, {Hasselquist}, {Hawley},
  {Hayden}, {Hearty}, {Hekker}, {Ho}, {Hogg}, {Holley-Bockelmann}, {Holtzman},
  {Honscheid}, {Huber}, {Huehnerhoff}, {Ivans}, {Jiang}, {Johnson},
  {Kinemuchi}, {Kirkby}, {Kitaura}, {Klaene}, {Knapp}, {Kneib}, {Koenig},
  {Lam}, {Lan}, {Lang}, {Laurent}, {Le Goff}, {Leauthaud}, {Lee}, {Lee},
  {Licquia}, {Liu}, {Long}, {L{\'o}pez-Corredoira}, {Lorenzo-Oliveira},
  {Lucatello}, {Lundgren}, {Lupton}, {Mack}, {Mahadevan}, {Maia}, {Majewski},
  {Malanushenko}, {Malanushenko}, {Manchado}, {Manera}, {Mao}, {Maraston},
  {Marchwinski}, {Margala}, {Martell}, {Martig}, {Masters}, {Mathur},
  {McBride}, {McGehee}, {McGreer}, {McMahon}, {M{\'e}nard}, {Menzel},
  {Merloni}, {M{\'e}sz{\'a}ros}, {Miller}, {Miralda-Escud{\'e}}, {Miyatake},
  {Montero-Dorta}, {More}, {Morganson}, {Morice-Atkinson}, {Morrison},
  {Mosser}, {Muna}, {Myers}, {Nandra}, {Newman}, {Neyrinck}, {Nguyen},
  {Nichol}, {Nidever}, {Noterdaeme}, {Nuza}, {O'Connell}, {O'Connell},
  {O'Connell}, {Ogando}, {Olmstead}, {Oravetz}, {Oravetz}, {Osumi}, {Owen},
  {Padgett}, {Padmanabhan}, {Paegert}, {Palanque-Delabrouille}, {Pan},
  {Parejko}, {P{\^a}ris}, {Park}, {Pattarakijwanich}, {Pellejero-Ibanez},
  {Pepper}, {Percival}, {P{\'e}rez-Fournon}, {P{\'e}rez-R{\`a}fols},
  {Petitjean}, {Pieri}, {Pinsonneault}, {Porto de Mello}, {Prada}, {Prakash},
  {Price-Whelan}, {Protopapas}, {Raddick}, {Rahman}, {Reid}, {Rich}, {Rix},
  {Robin}, {Rockosi}, {Rodrigues}, {Rodr{\'\i}guez-Torres}, {Roe}, {Ross},
  {Ross}, {Rossi}, {Ruan}, {Rubi{\~n}o-Mart{\'\i}n}, {Rykoff},
  {Salazar-Albornoz}, {Salvato}, {Samushia}, {S{\'a}nchez}, {Santiago},
  {Sayres}, {Schiavon}, {Schlegel}, {Schmidt}, {Schneider}, {Schultheis},
  {Schwope}, {Sc{\'o}ccola}, {Scott}, {Sellgren}, {Seo}, {Serenelli}, {Shane},
  {Shen}, {Shetrone}, {Shu}, {Silva Aguirre}, {Sivarani}, {Skrutskie},
  {Slosar}, {Smith}, {Sobreira}, {Souto}, {Stassun}, {Steinmetz}, {Stello},
  {Strauss}, {Streblyanska}, {Suzuki}, {Swanson}, {Tan}, {Tayar}, {Terrien},
  {Thakar}, {Thomas}, {Thomas}, {Thompson}, {Tinker}, {Tojeiro}, {Troup},
  {Vargas-Maga{\~n}a}, {Vazquez}, {Verde}, {Viel}, {Vogt}, {Wake}, {Wang},
  {Weaver}, {Weinberg}, {Weiner}, {White}, {Wilson}, {Wisniewski},
  {Wood-Vasey}, {Ye`che}, {York}, {Zakamska}, {Zamora}, {Zasowski}, {Zehavi},
  {Zhao}, {Zheng}, {Zhou}, {Zhou}, {Zou}, \& {Zhu}}]{Alam2015}
{Alam}, S., {Albareti}, F.~D., {Allende Prieto}, C., {et~al.} 2015, \apjs, 219,
  12, \dodoi{10.1088/0067-0049/219/1/12}

\bibitem[{{Alexander} {et~al.}(2022){Alexander}, {Capanelli}, {G.~M. Ferreira},
  \& {McDonough}}]{2022PhLB..83337298A}
{Alexander}, S., {Capanelli}, C., {G.~M. Ferreira}, E., \& {McDonough}, E.
  2022, Physics Letters B, 833, 137298, \dodoi{10.1016/j.physletb.2022.137298}

\bibitem[{{Aragon Calvo} {et~al.}(2019){Aragon Calvo}, {Neyrinck}, \&
  {Silk}}]{Aragon2016galaxyquenching}
{Aragon Calvo}, M.~A., {Neyrinck}, M.~C., \& {Silk}, J. 2019, The Open Journal
  of Astrophysics, 2, 7, \dodoi{10.21105/astro.1697.07881}

\bibitem[{{Aragon-Calvo} \& {Yang}(2014)}]{2014MNRAS.440L..46A}
{Aragon-Calvo}, M.~A., \& {Yang}, L.~F. 2014, \mnras, 440, L46,
  \dodoi{10.1093/mnrasl/slu009}

\bibitem[{{Bershady} {et~al.}(2024){Bershady}, {Westfall}, {Shetty}, {Law},
  {Cappellari}, {Drory}, {Bundy}, \& {Yan}}]{bershady2024asymmetric}
{Bershady}, M.~A., {Westfall}, K.~B., {Shetty}, S., {et~al.} 2024, \mnras, 531,
  1592, \dodoi{10.1093/mnras/stae1207}

\bibitem[{{Bond} {et~al.}(1996){Bond}, {Kofman}, \&
  {Pogosyan}}]{Bond1996Nature}
{Bond}, J.~R., {Kofman}, L., \& {Pogosyan}, D. 1996, \nat, 380, 603,
  \dodoi{10.1038/380603a0}

\bibitem[{{Bundy} {et~al.}(2015){Bundy}, {Bershady}, {Law}, {Yan}, {Drory},
  {MacDonald}, {Wake}, {Cherinka}, {S{\'a}nchez-Gallego}, {Weijmans}, {Thomas},
  {Tremonti}, {Masters}, {Coccato}, {Diamond-Stanic}, {Arag{\'o}n-Salamanca},
  {Avila-Reese}, {Badenes}, {Falc{\'o}n-Barroso}, {Belfiore}, {Bizyaev},
  {Blanc}, {Bland-Hawthorn}, {Blanton}, {Brownstein}, {Byler}, {Cappellari},
  {Conroy}, {Dutton}, {Emsellem}, {Etherington}, {Frinchaboy}, {Fu}, {Gunn},
  {Harding}, {Johnston}, {Kauffmann}, {Kinemuchi}, {Klaene}, {Knapen},
  {Leauthaud}, {Li}, {Lin}, {Maiolino}, {Malanushenko}, {Malanushenko}, {Mao},
  {Maraston}, {McDermid}, {Merrifield}, {Nichol}, {Oravetz}, {Pan}, {Parejko},
  {Sanchez}, {Schlegel}, {Simmons}, {Steele}, {Steinmetz}, {Thanjavur},
  {Thompson}, {Tinker}, {van den Bosch}, {Westfall}, {Wilkinson}, {Wright},
  {Xiao}, \& {Zhang}}]{MaNGA2015Bundy}
{Bundy}, K., {Bershady}, M.~A., {Law}, D.~R., {et~al.} 2015, \apj, 798, 7,
  \dodoi{10.1088/0004-637X/798/1/7}

\bibitem[{{Codis} {et~al.}(2012){Codis}, {Pichon}, {Devriendt}, {Slyz},
  {Pogosyan}, {Dubois}, \& {Sousbie}}]{Codis2012}
{Codis}, S., {Pichon}, C., {Devriendt}, J., {et~al.} 2012, \mnras, 427, 3320,
  \dodoi{10.1111/j.1365-2966.2012.21636.x}

\bibitem[{{Colless} {et~al.}(2003){Colless}, {Peterson}, {Jackson}, {Peacock},
  {Cole}, {Norberg}, {Baldry}, {Baugh}, {Bland-Hawthorn}, {Bridges}, {Cannon},
  {Collins}, {Couch}, {Cross}, {Dalton}, {De Propris}, {Driver}, {Efstathiou},
  {Ellis}, {Frenk}, {Glazebrook}, {Lahav}, {Lewis}, {Lumsden}, {Maddox},
  {Madgwick}, {Sutherland}, \& {Taylor}}]{Colless2003}
{Colless}, M., {Peterson}, B.~A., {Jackson}, C., {et~al.} 2003, arXiv e-prints,
  astro, \dodoi{10.48550/arXiv.astro-ph/0306581}

\bibitem[{{Davis} {et~al.}(1985){Davis}, {Efstathiou}, {Frenk}, \&
  {White}}]{FoF1985}
{Davis}, M., {Efstathiou}, G., {Frenk}, C.~S., \& {White}, S.~D.~M. 1985, \apj,
  292, 371, \dodoi{10.1086/163168}

\bibitem[{{Dolag} {et~al.}(2009){Dolag}, {Borgani}, {Murante}, \&
  {Springel}}]{subfindDolag2009}
{Dolag}, K., {Borgani}, S., {Murante}, G., \& {Springel}, V. 2009, \mnras, 399,
  497, \dodoi{10.1111/j.1365-2966.2009.15034.x}

\bibitem[{{Dubois} {et~al.}(2014){Dubois}, {Pichon}, {Welker}, {Le Borgne},
  {Devriendt}, {Laigle}, {Codis}, {Pogosyan}, {Arnouts}, {Benabed}, {Bertin},
  {Blaizot}, {Bouchet}, {Cardoso}, {Colombi}, {de Lapparent}, {Desjacques},
  {Gavazzi}, {Kassin}, {Kimm}, {McCracken}, {Milliard}, {Peirani}, {Prunet},
  {Rouberol}, {Silk}, {Slyz}, {Sousbie}, {Teyssier}, {Tresse}, {Treyer},
  {Vibert}, \& {Volonteri}}]{dubois2014dancing}
{Dubois}, Y., {Pichon}, C., {Welker}, C., {et~al.} 2014, \mnras, 444, 1453,
  \dodoi{10.1093/mnras/stu1227}

\bibitem[{{Gal{\'a}rraga-Espinosa} {et~al.}(2020){Gal{\'a}rraga-Espinosa},
  {Aghanim}, {Langer}, {Gouin}, \& {Malavasi}}]{Daniela2020}
{Gal{\'a}rraga-Espinosa}, D., {Aghanim}, N., {Langer}, M., {Gouin}, C., \&
  {Malavasi}, N. 2020, \aap, 641, A173, \dodoi{10.1051/0004-6361/202037986}

\bibitem[{{Gal{\'a}rraga-Espinosa} {et~al.}(2022){Gal{\'a}rraga-Espinosa},
  {Langer}, \& {Aghanim}}]{Daniela2022}
{Gal{\'a}rraga-Espinosa}, D., {Langer}, M., \& {Aghanim}, N. 2022, \aap, 661,
  A115, \dodoi{10.1051/0004-6361/202141974}

\bibitem[{{Gal{\'a}rraga-Espinosa} {et~al.}(2024){Gal{\'a}rraga-Espinosa},
  {Cadiou}, {Gouin}, {White}, {Springel}, {Pakmor}, {Hadzhiyska}, {Bose},
  {Ferlito}, {Hernquist}, {Kannan}, {Barrera}, {Maria Delgado}, \&
  {Hern{\'a}ndez-Aguayo}}]{Daniela2024}
{Gal{\'a}rraga-Espinosa}, D., {Cadiou}, C., {Gouin}, C., {et~al.} 2024, \aap,
  684, A63, \dodoi{10.1051/0004-6361/202347982}

\bibitem[{{Ganeshaiah Veena} {et~al.}(2019){Ganeshaiah Veena}, {Cautun},
  {Tempel}, {van de Weygaert}, \& {Frenk}}]{ganeshaiah2019cosmic}
{Ganeshaiah Veena}, P., {Cautun}, M., {Tempel}, E., {van de Weygaert}, R., \&
  {Frenk}, C.~S. 2019, \mnras, 487, 1607, \dodoi{10.1093/mnras/stz1343}

\bibitem[{{Ganeshaiah Veena} {et~al.}(2021){Ganeshaiah Veena}, {Cautun}, {van
  de Weygaert}, {Tempel}, \& {Frenk}}]{ganeshaiah2021cosmic}
{Ganeshaiah Veena}, P., {Cautun}, M., {van de Weygaert}, R., {Tempel}, E., \&
  {Frenk}, C.~S. 2021, \mnras, 503, 2280, \dodoi{10.1093/mnras/stab411}

\bibitem[{{Ganeshaiah Veena} {et~al.}(2018){Ganeshaiah Veena}, {Cautun}, {van
  de Weygaert}, {Tempel}, {Jones}, {Rieder}, \& {Frenk}}]{ganeshaiah2018cosmic}
{Ganeshaiah Veena}, P., {Cautun}, M., {van de Weygaert}, R., {et~al.} 2018,
  \mnras, 481, 414, \dodoi{10.1093/mnras/sty2270}

\bibitem[{{Gregory} \& {Thompson}(1978)}]{Gregory1978}
{Gregory}, S.~A., \& {Thompson}, L.~A. 1978, \apj, 222, 784,
  \dodoi{10.1086/156198}

\bibitem[{{Hahn} {et~al.}(2007){Hahn}, {Carollo}, {Porciani}, \&
  {Dekel}}]{2007MNRAS.381...41H}
{Hahn}, O., {Carollo}, C.~M., {Porciani}, C., \& {Dekel}, A. 2007, \mnras, 381,
  41, \dodoi{10.1111/j.1365-2966.2007.12249.x}

\bibitem[{{Huchra} {et~al.}(2005){Huchra}, {Jarrett}, {Skrutskie}, {Cutri},
  {Schneider}, {Macri}, {Steining}, {Mader}, {Martimbeau}, \&
  {George}}]{Huchra2005}
{Huchra}, J., {Jarrett}, T., {Skrutskie}, M., {et~al.} 2005, 329, 135

\bibitem[{{Joeveer} \& {Einasto}(1978)}]{Joeveer1978}
{Joeveer}, M., \& {Einasto}, J. 1978, 79, 241

\bibitem[{{Kang} \& {Wang}(2015)}]{KangWang2015}
{Kang}, X., \& {Wang}, P. 2015, \apj, 813, 6, \dodoi{10.1088/0004-637X/813/1/6}

\bibitem[{{Laigle} {et~al.}(2015){Laigle}, {Pichon}, {Codis}, {Dubois}, {Le
  Borgne}, {Pogosyan}, {Devriendt}, {Peirani}, {Prunet}, {Rouberol}, {Slyz}, \&
  {Sousbie}}]{Laigle2015}
{Laigle}, C., {Pichon}, C., {Codis}, S., {et~al.} 2015, \mnras, 446, 2744,
  \dodoi{10.1093/mnras/stu2289}

\bibitem[{{Lee} \& {Erdogdu}(2007)}]{Lee2007}
{Lee}, J., \& {Erdogdu}, P. 2007, \apj, 671, 1248, \dodoi{10.1086/523351}

\bibitem[{{Lee} \& {Libeskind}(2020)}]{2020ApJ...902...22L}
{Lee}, J., \& {Libeskind}, N.~I. 2020, \apj, 902, 22,
  \dodoi{10.3847/1538-4357/abb314}

\bibitem[{{Lee} {et~al.}(2020){Lee}, {Libeskind}, \&
  {Ryu}}]{2020ApJ...898L..27L}
{Lee}, J., {Libeskind}, N.~I., \& {Ryu}, S. 2020, \apjl, 898, L27,
  \dodoi{10.3847/2041-8213/aba2ee}

\bibitem[{{Libeskind} {et~al.}(2013){Libeskind}, {Hoffman}, {Forero-Romero},
  {Gottl{\"o}ber}, {Knebe}, {Steinmetz}, \& {Klypin}}]{Libeskind2013}
{Libeskind}, N.~I., {Hoffman}, Y., {Forero-Romero}, J., {et~al.} 2013, \mnras,
  428, 2489, \dodoi{10.1093/mnras/sts216}

\bibitem[{{Libeskind} {et~al.}(2014){Libeskind}, {Knebe}, {Hoffman}, \&
  {Gottl{\"o}ber}}]{Libeskind2014}
{Libeskind}, N.~I., {Knebe}, A., {Hoffman}, Y., \& {Gottl{\"o}ber}, S. 2014,
  \mnras, 443, 1274, \dodoi{10.1093/mnras/stu1216}

\bibitem[{{Libeskind} {et~al.}(2015){Libeskind}, {Tempel}, {Hoffman}, {Tully},
  \& {Courtois}}]{Libeskind2015bisous}
{Libeskind}, N.~I., {Tempel}, E., {Hoffman}, Y., {Tully}, R.~B., \& {Courtois},
  H. 2015, \mnras, 453, L108, \dodoi{10.1093/mnrasl/slv099}

\bibitem[{{Libeskind} {et~al.}(2018){Libeskind}, {van de Weygaert}, {Cautun},
  {Falck}, {Tempel}, {Abel}, {Alpaslan}, {Arag{\'o}n-Calvo}, {Forero-Romero},
  {Gonzalez}, {Gottl{\"o}ber}, {Hahn}, {Hellwing}, {Hoffman}, {Jones},
  {Kitaura}, {Knebe}, {Manti}, {Neyrinck}, {Nuza}, {Padilla}, {Platen},
  {Ramachandra}, {Robotham}, {Saar}, {Shandarin}, {Steinmetz}, {Stoica},
  {Sousbie}, \& {Yepes}}]{Libeskind2018}
{Libeskind}, N.~I., {van de Weygaert}, R., {Cautun}, M., {et~al.} 2018, \mnras,
  473, 1195, \dodoi{10.1093/mnras/stx1976}

\bibitem[{{Marinacci} {et~al.}(2018){Marinacci}, {Vogelsberger}, {Pakmor},
  {Torrey}, {Springel}, {Hernquist}, {Nelson}, {Weinberger}, {Pillepich},
  {Naiman}, \& {Genel}}]{TNGpaper1}
{Marinacci}, F., {Vogelsberger}, M., {Pakmor}, R., {et~al.} 2018, \mnras, 480,
  5113, \dodoi{10.1093/mnras/sty2206}

\bibitem[{{Moon} \& {Lee}(2023)}]{2023ApJ...945...13M}
{Moon}, J.-S., \& {Lee}, J. 2023, \apj, 945, 13,
  \dodoi{10.3847/1538-4357/acac8e}

\bibitem[{{Moon} \& {Lee}(2024)}]{2024ApJ...966..100M}
---. 2024, \apj, 966, 100, \dodoi{10.3847/1538-4357/ad3825}

\bibitem[{{Naiman} {et~al.}(2018){Naiman}, {Pillepich}, {Springel},
  {Ramirez-Ruiz}, {Torrey}, {Vogelsberger}, {Pakmor}, {Nelson}, {Marinacci},
  {Hernquist}, {Weinberger}, \& {Genel}}]{TNGpaper2}
{Naiman}, J.~P., {Pillepich}, A., {Springel}, V., {et~al.} 2018, \mnras, 477,
  1206, \dodoi{10.1093/mnras/sty618}

\bibitem[{{Neyrinck} {et~al.}(2020){Neyrinck}, {Aragon-Calvo}, {Falck},
  {Szalay}, \& {Wang}}]{Neyrinck2020}
{Neyrinck}, M., {Aragon-Calvo}, M.~A., {Falck}, B., {Szalay}, A.~S., \& {Wang},
  J. 2020, The Open Journal of Astrophysics, 3, 3,
  \dodoi{10.21105/astro.1904.03201}

\bibitem[{{Pahwa} {et~al.}(2016){Pahwa}, {Libeskind}, {Tempel}, {Hoffman},
  {Tully}, {Courtois}, {Gottl{\"o}ber}, {Steinmetz}, \& {Sorce}}]{Pahwa2016}
{Pahwa}, I., {Libeskind}, N.~I., {Tempel}, E., {et~al.} 2016, \mnras, 457, 695,
  \dodoi{10.1093/mnras/stv2930}

\bibitem[{{Pillepich} {et~al.}(2018){Pillepich}, {Nelson}, {Hernquist},
  {Springel}, {Pakmor}, {Torrey}, {Weinberger}, {Genel}, {Naiman}, {Marinacci},
  \& {Vogelsberger}}]{TNGpaper4}
{Pillepich}, A., {Nelson}, D., {Hernquist}, L., {et~al.} 2018, \mnras, 475,
  648, \dodoi{10.1093/mnras/stx3112}

\bibitem[{{Planck Collaboration} {et~al.}(2016){Planck Collaboration}, {Ade},
  {Aghanim}, {Arnaud}, {Ashdown}, {Aumont}, {Baccigalupi}, {Banday},
  {Barreiro}, {Bartlett}, {Bartolo}, {Battaner}, {Battye}, {Benabed},
  {Beno{\^\i}t}, {Benoit-L{\'e}vy}, {Bernard}, {Bersanelli}, {Bielewicz},
  {Bock}, {Bonaldi}, {Bonavera}, {Bond}, {Borrill}, {Bouchet}, {Boulanger},
  {Bucher}, {Burigana}, {Butler}, {Calabrese}, {Cardoso}, {Catalano},
  {Challinor}, {Chamballu}, {Chary}, {Chiang}, {Chluba}, {Christensen},
  {Church}, {Clements}, {Colombi}, {Colombo}, {Combet}, {Coulais}, {Crill},
  {Curto}, {Cuttaia}, {Danese}, {Davies}, {Davis}, {de Bernardis}, {de Rosa},
  {de Zotti}, {Delabrouille}, {D{\'e}sert}, {Di Valentino}, {Dickinson},
  {Diego}, {Dolag}, {Dole}, {Donzelli}, {Dor{\'e}}, {Douspis}, {Ducout},
  {Dunkley}, {Dupac}, {Efstathiou}, {Elsner}, {En{\ss}lin}, {Eriksen},
  {Farhang}, {Fergusson}, {Finelli}, {Forni}, {Frailis}, {Fraisse},
  {Franceschi}, {Frejsel}, {Galeotta}, {Galli}, {Ganga}, {Gauthier}, {Gerbino},
  {Ghosh}, {Giard}, {Giraud-H{\'e}raud}, {Giusarma}, {Gjerl{\o}w},
  {Gonz{\'a}lez-Nuevo}, {G{\'o}rski}, {Gratton}, {Gregorio}, {Gruppuso},
  {Gudmundsson}, {Hamann}, {Hansen}, {Hanson}, {Harrison}, {Helou},
  {Henrot-Versill{\'e}}, {Hern{\'a}ndez-Monteagudo}, {Herranz}, {Hildebrandt},
  {Hivon}, {Hobson}, {Holmes}, {Hornstrup}, {Hovest}, {Huang}, {Huffenberger},
  {Hurier}, {Jaffe}, {Jaffe}, {Jones}, {Juvela}, {Keih{\"a}nen}, {Keskitalo},
  {Kisner}, {Kneissl}, {Knoche}, {Knox}, {Kunz}, {Kurki-Suonio}, {Lagache},
  {L{\"a}hteenm{\"a}ki}, {Lamarre}, {Lasenby}, {Lattanzi}, {Lawrence}, {Leahy},
  {Leonardi}, {Lesgourgues}, {Levrier}, {Lewis}, {Liguori}, {Lilje},
  {Linden-V{\o}rnle}, {L{\'o}pez-Caniego}, {Lubin}, {Mac{\'\i}as-P{\'e}rez},
  {Maggio}, {Maino}, {Mandolesi}, {Mangilli}, {Marchini}, {Maris}, {Martin},
  {Martinelli}, {Mart{\'\i}nez-Gonz{\'a}lez}, {Masi}, {Matarrese}, {McGehee},
  {Meinhold}, {Melchiorri}, {Melin}, {Mendes}, {Mennella}, {Migliaccio},
  {Millea}, {Mitra}, {Miville-Desch{\^e}nes}, {Moneti}, {Montier}, {Morgante},
  {Mortlock}, {Moss}, {Munshi}, {Murphy}, {Naselsky}, {Nati}, {Natoli},
  {Netterfield}, {N{\o}rgaard-Nielsen}, {Noviello}, {Novikov}, {Novikov},
  {Oxborrow}, {Paci}, {Pagano}, {Pajot}, {Paladini}, {Paoletti}, {Partridge},
  {Pasian}, {Patanchon}, {Pearson}, {Perdereau}, {Perotto}, {Perrotta},
  {Pettorino}, {Piacentini}, {Piat}, {Pierpaoli}, {Pietrobon}, {Plaszczynski},
  {Pointecouteau}, {Polenta}, {Popa}, {Pratt}, {Pr{\'e}zeau}, {Prunet},
  {Puget}, {Rachen}, {Reach}, {Rebolo}, {Reinecke}, {Remazeilles}, {Renault},
  {Renzi}, {Ristorcelli}, {Rocha}, {Rosset}, {Rossetti}, {Roudier},
  {Rouill{\'e} d'Orfeuil}, {Rowan-Robinson}, {Rubi{\~n}o-Mart{\'\i}n},
  {Rusholme}, {Said}, {Salvatelli}, {Salvati}, {Sandri}, {Santos},
  {Savelainen}, {Savini}, {Scott}, {Seiffert}, {Serra}, {Shellard}, {Spencer},
  {Spinelli}, {Stolyarov}, {Stompor}, {Sudiwala}, {Sunyaev}, {Sutton},
  {Suur-Uski}, {Sygnet}, {Tauber}, {Terenzi}, {Toffolatti}, {Tomasi},
  {Tristram}, {Trombetti}, {Tucci}, {Tuovinen}, {T{\"u}rler}, {Umana},
  {Valenziano}, {Valiviita}, {Van Tent}, {Vielva}, {Villa}, {Wade}, {Wandelt},
  {Wehus}, {White}, {White}, {Wilkinson}, {Yvon}, {Zacchei}, \&
  {Zonca}}]{Planck2016}
{Planck Collaboration}, {Ade}, P.~A.~R., {Aghanim}, N., {et~al.} 2016, \aap,
  594, A13, \dodoi{10.1051/0004-6361/201525830}

\bibitem[{{Shectman} {et~al.}(1996){Shectman}, {Landy}, {Oemler}, {Tucker},
  {Lin}, {Kirshner}, \& {Schechter}}]{Shectman1996}
{Shectman}, S.~A., {Landy}, S.~D., {Oemler}, A., {et~al.} 1996, \apj, 470, 172,
  \dodoi{10.1086/177858}

\bibitem[{{Sheng} {et~al.}(2022){Sheng}, {Li}, {Yu}, {Wang}, {Wang}, \&
  {Kang}}]{Sheng2022PhRvD}
{Sheng}, M.-J., {Li}, S., {Yu}, H.-R., {et~al.} 2022, \prd, 105, 063540,
  \dodoi{10.1103/PhysRevD.105.063540}

\bibitem[{{Shi} {et~al.}(2015){Shi}, {Wang}, \& {Mo}}]{ShiJingJing2015}
{Shi}, J., {Wang}, H., \& {Mo}, H.~J. 2015, \apj, 807, 37,
  \dodoi{10.1088/0004-637X/807/1/37}

\bibitem[{{Springel} {et~al.}(2001){Springel}, {White}, {Tormen}, \&
  {Kauffmann}}]{sunfindSpringel2001}
{Springel}, V., {White}, S. D.~M., {Tormen}, G., \& {Kauffmann}, G. 2001,
  \mnras, 328, 726, \dodoi{10.1046/j.1365-8711.2001.04912.x}

\bibitem[{{Springel} {et~al.}(2018){Springel}, {Pakmor}, {Pillepich},
  {Weinberger}, {Nelson}, {Hernquist}, {Vogelsberger}, {Genel}, {Torrey},
  {Marinacci}, \& {Naiman}}]{TNGpaper5}
{Springel}, V., {Pakmor}, R., {Pillepich}, A., {et~al.} 2018, \mnras, 475, 676,
  \dodoi{10.1093/mnras/stx3304}

\bibitem[{{Tang} {et~al.}(2025){Tang}, {Wang}, {Wang}, {Sheng}, {Yu}, \&
  {Xu}}]{tangxx2025}
{Tang}, X.-x., {Wang}, P., {Wang}, W., {et~al.} 2025, \apj, 982, 197,
  \dodoi{10.3847/1538-4357/adbbd7}

\bibitem[{{Tegmark} {et~al.}(2004){Tegmark}, {Blanton}, {Strauss}, {Hoyle},
  {Schlegel}, {Scoccimarro}, {Vogeley}, {Weinberg}, {Zehavi}, {Berlind},
  {Budavari}, {Connolly}, {Eisenstein}, {Finkbeiner}, {Frieman}, {Gunn},
  {Hamilton}, {Hui}, {Jain}, {Johnston}, {Kent}, {Lin}, {Nakajima}, {Nichol},
  {Ostriker}, {Pope}, {Scranton}, {Seljak}, {Sheth}, {Stebbins}, {Szalay},
  {Szapudi}, {Verde}, {Xu}, {Annis}, {Bahcall}, {Brinkmann}, {Burles},
  {Castander}, {Csabai}, {Loveday}, {Doi}, {Fukugita}, {Gott}, {Hennessy},
  {Hogg}, {Ivezi{\'c}}, {Knapp}, {Lamb}, {Lee}, {Lupton}, {McKay}, {Kunszt},
  {Munn}, {O'Connell}, {Peoples}, {Pier}, {Richmond}, {Rockosi}, {Schneider},
  {Stoughton}, {Tucker}, {Vanden Berk}, {Yanny}, {York}, \& {SDSS
  Collaboration}}]{Tegmark2004}
{Tegmark}, M., {Blanton}, M.~R., {Strauss}, M.~A., {et~al.} 2004, \apj, 606,
  702, \dodoi{10.1086/382125}

\bibitem[{{Tempel} \& {Libeskind}(2013)}]{TempelLibeskind2013}
{Tempel}, E., \& {Libeskind}, N.~I. 2013, \apjl, 775, L42,
  \dodoi{10.1088/2041-8205/775/2/L42}

\bibitem[{Tempel {et~al.}(2014)Tempel, Stoica, \& Martínez}]{Tempel2014bisous}
Tempel, E., Stoica, R.~S., \& Martínez, V.~J. 2014, Monthly Notices of the
  Royal Astronomical Society, 438, 3465, \dodoi{10.1093/mnras/stt2454}

\bibitem[{{Tempel} {et~al.}(2017){Tempel}, {Tuvikene}, {Kipper}, \&
  {Libeskind}}]{Tempel2017catalogue}
{Tempel}, E., {Tuvikene}, T., {Kipper}, R., \& {Libeskind}, N.~I. 2017, \aap,
  602, A100, \dodoi{10.1051/0004-6361/201730499}

\bibitem[{{Trowland} {et~al.}(2013){Trowland}, {Lewis}, \&
  {Bland-Hawthorn}}]{trowland2012cosmic}
{Trowland}, H.~E., {Lewis}, G.~F., \& {Bland-Hawthorn}, J. 2013, \apj, 762, 72,
  \dodoi{10.1088/0004-637X/762/2/72}

\bibitem[{{Trujillo} {et~al.}(2006){Trujillo}, {Carretero}, \&
  {Patiri}}]{Trujillo2006}
{Trujillo}, I., {Carretero}, C., \& {Patiri}, S.~G. 2006, \apjl, 640, L111,
  \dodoi{10.1086/503548}

\bibitem[{{Varela} {et~al.}(2012){Varela}, {Betancort-Rijo}, {Trujillo}, \&
  {Ricciardelli}}]{Varela2012}
{Varela}, J., {Betancort-Rijo}, J., {Trujillo}, I., \& {Ricciardelli}, E. 2012,
  \apj, 744, 82, \dodoi{10.1088/0004-637X/744/2/82}

\bibitem[{Wang {et~al.}(2020)Wang, Bose, Frenk, Gao, Jenkins, Springel, \&
  White}]{Wangjie2020}
Wang, J., Bose, S., Frenk, C.~S., {et~al.} 2020, Nature, 585, 39,
  \dodoi{10.1038/s41586-020-2642-9}

\bibitem[{{Wang} {et~al.}(2018){Wang}, {Guo}, {Kang}, \&
  {Libeskind}}]{Wang2018ApJ.spin}
{Wang}, P., {Guo}, Q., {Kang}, X., \& {Libeskind}, N.~I. 2018, \apj, 866, 138,
  \dodoi{10.3847/1538-4357/aae20f}

\bibitem[{{Wang} \& {Kang}(2017)}]{WangKang2017}
{Wang}, P., \& {Kang}, X. 2017, \mnras, 468, L123,
  \dodoi{10.1093/mnrasl/slx038}

\bibitem[{Wang \& Kang(2018)}]{WangKang2018}
Wang, P., \& Kang, X. 2018, Monthly Notices of the Royal Astronomical Society,
  473, 1562–1569, \dodoi{10.1093/mnras/stx2466}

\bibitem[{{Wang} {et~al.}(2021){Wang}, {Libeskind}, {Tempel}, {Kang}, \&
  {Guo}}]{Wang2021NatAS}
{Wang}, P., {Libeskind}, N.~I., {Tempel}, E., {Kang}, X., \& {Guo}, Q. 2021,
  Nature Astronomy, 5, 839, \dodoi{10.1038/s41550-021-01380-6}

\bibitem[{{Wang} {et~al.}(2024){Wang}, {Wang}, {Guo}, {Kang}, {Libeskind},
  {Gal{\'a}rraga-Espinosa}, {Springel}, {Kannan}, {Hernquist}, {Pakmor}, {Yu},
  {Bose}, {Guo}, {Yu}, \& {Hern{\'a}ndez-Aguayo}}]{2024MNRAS.532.4604W}
{Wang}, W., {Wang}, P., {Guo}, H., {et~al.} 2024, \mnras, 532, 4604,
  \dodoi{10.1093/mnras/stae1801}

\bibitem[{{Winkel} {et~al.}(2021){Winkel}, {Pasquali}, {Kraljic}, {Smith},
  {Gallazzi}, \& {Jackson}}]{winkel2021galaxyquenching}
{Winkel}, N., {Pasquali}, A., {Kraljic}, K., {et~al.} 2021, \mnras, 505, 4920,
  \dodoi{10.1093/mnras/stab1562}

\bibitem[{{Xia} {et~al.}(2021){Xia}, {Neyrinck}, {Cai}, \&
  {Arag{\'o}n-Calvo}}]{xia2021filaspin}
{Xia}, Q., {Neyrinck}, M.~C., {Cai}, Y.-C., \& {Arag{\'o}n-Calvo}, M.~A. 2021,
  \mnras, 506, 1059, \dodoi{10.1093/mnras/stab1713}

\bibitem[{{Zel'dovich}(1970)}]{Zeldovich1970}
{Zel'dovich}, Y.~B. 1970, \aap, 5, 84

\bibitem[{{Zhang} {et~al.}(2009){Zhang}, {Yang}, {Faltenbacher}, {Springel},
  {Lin}, \& {Wang}}]{Zhang2009ApJ}
{Zhang}, Y., {Yang}, X., {Faltenbacher}, A., {et~al.} 2009, \apj, 706, 747,
  \dodoi{10.1088/0004-637X/706/1/747}

\bibitem[{{Zhang} {et~al.}(2015){Zhang}, {Yang}, {Wang}, {Wang}, {Luo}, {Mo},
  \& {van den Bosch}}]{Zhang2015ApJ}
{Zhang}, Y., {Yang}, X., {Wang}, H., {et~al.} 2015, \apj, 798, 17,
  \dodoi{10.1088/0004-637X/798/1/17}

\bibitem[{{Zheng} {et~al.}(2023){Zheng}, {Cai}, {Zhu}, {Neyrinck}, {Wang}, \&
  {Li}}]{2023MNRAS.519.1171Z}
{Zheng}, Y., {Cai}, Y.-C., {Zhu}, W., {et~al.} 2023, \mnras, 519, 1171,
  \dodoi{10.1093/mnras/stac3600}

\end{thebibliography}
\bibliographystyle{aasjournal}




 \end{CJK*}
\end{document}